# Fabrication methods for integrating 2D materials

*David J. Moss*

Optical Sciences Centre, Swinburne University of Technology, Hawthorn, VIC 3122, Australia

## Abstract

With compact footprint, low energy consumption, high scalability, and mass producibility, chip-scale integrated devices are an indispensable part of modern technological change and development. Recent advances in two-dimensional (2D) layered materials with their unique structures and distinctive properties have motivated their on-chip integration, yielding a variety of functional devices with superior performance and new features. To realize integrated devices incorporating 2D materials, it requires a diverse range of device fabrication techniques, which are of fundamental importance to achieve good performance and high reproducibility. This paper reviews the state-of-art fabrication techniques for the on-chip integration of 2D materials. First, an overview of the material properties and on-chip applications of 2D materials is provided. Second, different approaches used for integrating 2D materials on chips are comprehensively reviewed, which are categorized into material synthesis, on-chip transfer, film patterning, and property tuning / modification. Third, the methods for integrating 2D van der Waals heterostructures are also discussed and summarized. Finally, the current challenges and future perspectives are highlighted.

## 1. Introduction

The integrated circuits industry (both electronic and photonic) is fundamental to our information age. Chip-scale integrated devices, featuring compact footprint, low energy consumption, and low cost enabled by large-scale manufacturing, have had a profound influence on our modern life. Although complementary metal-oxide-semiconductor (CMOS) compatible material platforms, such as silicon, silicon nitride, and silica, have dominated integrated devices, they suffer from limitations intrinsically arising from their material properties, creating challenges for them to meet the ever-increasing demands for device functionality and performance.[1-3] On-chip integration of other materials has proven to be an attractive solution to overcome these challenges. For example, integrated spintronic transistors have been realized by integrating europium monoxide (EuO) ferromagnetic films to inject spin-polarized carriers into silicon.[4] In addition, the integration of organic polymers onto silicon photonic chips has greatly increased their processing speed and functionality, which were otherwise limited by the slow free carrier response.[5]

Since the ground-breaking discovery of graphene in 2004,[6] 2D layered materials have attracted tremendous interest, with the material family growing rapidly to include graphene oxide (GO), transition metal dichalcogenides (TMDCs), black phosphorus (BP), hexagonal boron nitride (h-BN), and many others.[7-11] Compared with bulk materials, the 2D counterparts exhibit many extraordinary properties, such as ultra-high carrier mobility, layer-dependent bandgaps, high anisotropy, broadband and low optical dispersion, and excellent nonlinear optical responses.[8-13] In addition, due to the weak out-of-plane van der Waals interactions, 2D materials possess surfaces that are free of dangling bonds, thus enabling them to be readily integrated onto chips without stress or restrictions due to lattice mismatch.[14, 15] Their atomically thin geometry is also advantageous for high-density integration and low-power operation. Together with their high degree of compatibility with integrated devices, these materials bring about exciting new opportunities for realizing novel functions and improved performance beyond what the conventional integrated platforms can possibly offer.

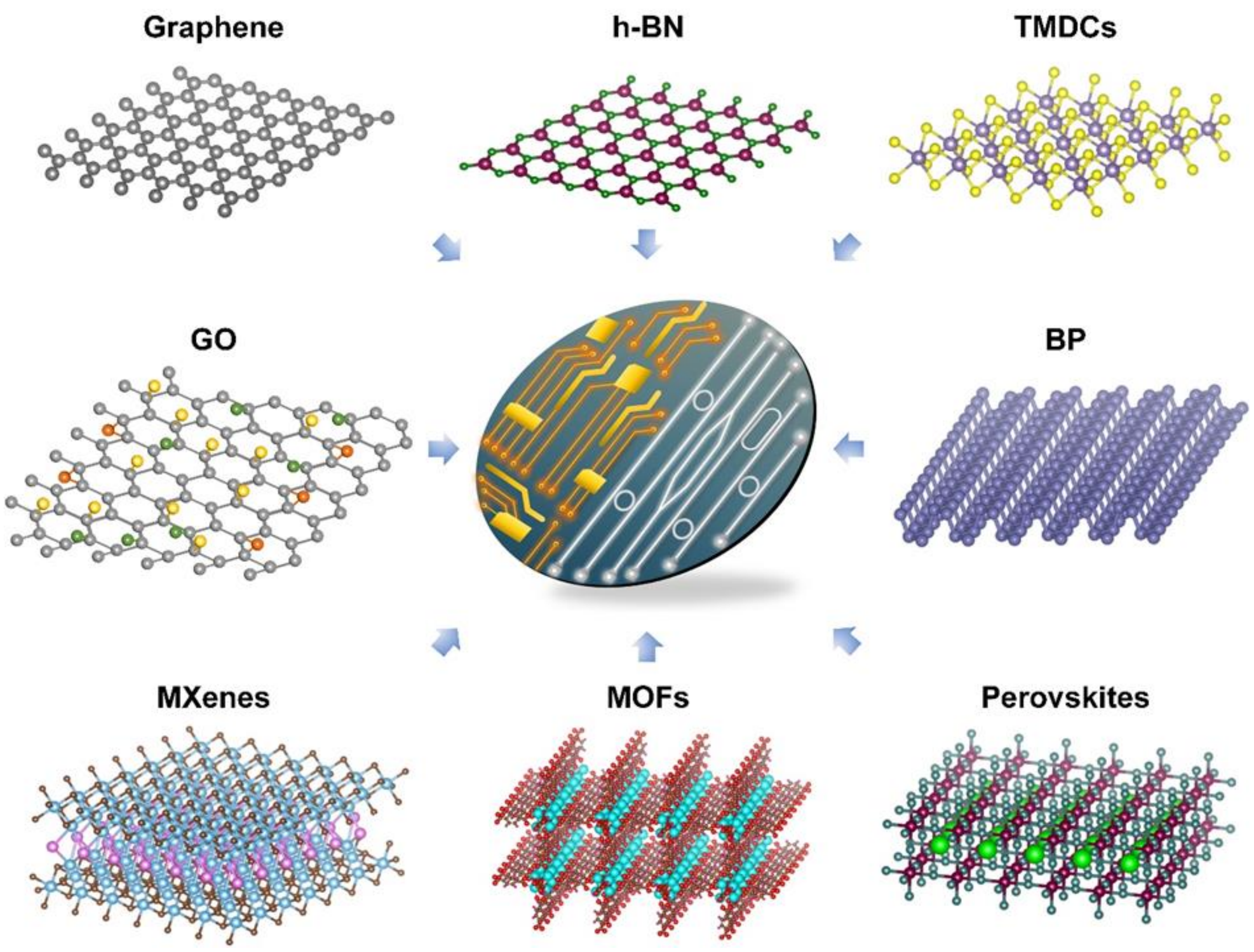

**Figure 1.** Illustration of typical 2D layered materials for on-chip integration. h-BN: hexagonal boron nitride. TMDCs: transition metal dichalcogenides. GO: graphene oxide. BP: black phosphorus. MOFs: metal-organic frameworks.

The incorporation of 2D materials onto conventional integrated circuits combines the best of both worlds (**Fig. 1**). In the past decade, there has been significant progress in hybrid integrated devices incorporating 2D materials for a variety of applications such as field-effect transistors,[16, 17] logic gate circuits,[18, 19] laser sources,[20, 21] optical modulators,[22, 23] photodetectors,[24, 25] polarizers,[26, 27] nonlinear optical devices,[28, 29] and sensors.[30-32] The success of these devices has relied heavily on developing advanced integration fabrication techniques – from material synthesis to on-chip transfer, film patterning, and property tuning / modification. While 2D materials have been the subject of many previous reviews,[14, 33-36] these have predominantly focused on either the material aspects of 2D films or the specific applications, rather than the fabrication techniques. Here, we focus on the latter – review comprehensively the diverse methods used to fabricate chip-scale integrated devices incorporating 2D materials, highlighting the challenges and potential for large-scale production in industrial settings outside of the laboratory.

This review is structured as follows. In Section 2, the properties of different 2D materials are briefly introduced, together with a summary of their on-chip applications. Next, the fabrication techniques for the on-chip integration of 2D materials along with their advantages and drawbacks are summarized and discussed in Section 3, being categorized into material synthesis, on-chip transfer, film patterning, and property tuning / modification. In Section 4, we review and discuss emerging techniques to implement 2D van der Waals heterostructures on chips. The current challenges and future perspectives are discussed in Section 5, followed by conclusions in Section 6.

## 2. 2D materials and their on-chip applications

The unique physics of atomic dimensional 2D materials give rise to remarkable properties compared to their bulk counterparts,[14, 37] and many have been successfully synthesized and incorporated onto integrated chips. In this section, we briefly introduce the properties and on-chip applications of 2D materials such as graphene, GO, TMDCs, BP, and h-BN, and other emerging materials.

### 2.1. Material properties

**Table 1** compares the basic properties of different 2D materials, including the bandgap, carrier mobility, photoluminescence (PL) wavelength, and their linear and nonlinear optical (NLO) properties. As seen in **Table 1**, these materials offer their own unique advantages for different applications.

**Table 1.** Comparison of basic material property parameters of some typical 2D materials. PL: photoluminescence. LO: linear optical. NLO: nonlinear optical.

| Material | Bandgap (eV) | Carrier mobility[a)] ($cm^2$ $V^{-1}s^{-1}$) | PL wavelength (nm) | LO parameters[b)] | | NLO parameters[c)] | | Ref. |
|---|---|---|---|---|---|---|---|---|
| | | | | *n* | *k* | $n_2$ ($m^2 \cdot W^{-1}$) | $\beta$ ($m \cdot W^{-1}$) | |
| Graphene | 0 | $1.2\times10^{5}$ | 354–729 | 2.70 | 1.35 | $-8.0\times10^{-14}$ | $9.0\times10^{-8}$ | [38-42] |
| GO | 2.4 | — | 430–1100 | 1.97 | < 0.01 | $1.5\times10^{-14}$ | $-5.3\times10^{-8}$ | [43-46] |
| $MoS_2$ | 1.2–1.8 | $2.0\times10^{2}$ | 670 | 3.65 | 0 | $1.8\times10^{-16}$ | $-3.8\times10^{-11}$ | [47-50] |
| $WSe_2$ | 1.2–1.7 | $2.5\times10^{2}$ | 750 | 3.70 | 0.10 | $-1.7\times10^{-15}$ | $2.1\times10^{-8}$ | [51-54] |
| $PdSe_2$ | 0–1.3 | $1.6\times10^{2}$ | — | 3.74 | 0.26 | $-1.3\times10^{-15}$ | $3.2\times10^{-8}$ | [55, 56] |
| $PtSe_2$ | 0–1.2 | $2.1\times10^{2}$ | — | 2.80 | 0.25 | $-3.7\times10^{-15}$ | $-8.8\times10^{-8}$ | [57-60] |
| SnSe | 0.86–1.63 | $10^{3}$ | 714 | 2.50 | 0.10 | — | $1.9\times10^{-9}$ | [61-65] |
| BP | 0.3–2.0 | $10^{3}$ | 918 | 2.90 | 0 | $6.8\times10^{-13}$ | $4.5\times10^{-10}$ | [66-70] |
| h-BN | 6.0 | — | 203 | 2.17 | — | $1.2\times10^{-12}$ | $5.0\times10^{-7}$ | [71-74] |
| $MoO_3$ | 3.3 | $3.0\times10^{3}$ | 414–670 | 2.36 | < 0.01 | — | $-1.7\times10^{-10}$ | [75-79] |
| $Bi_2Te_3$ | 0.16–1.36 | $2.5\times10^{2}$ | 621 | 5.60 | 2.10 | $-4.4\times10^{-10}$ | $2.3\times10^{-5}$ | [80-84] |
| $Nb_2C$ | 0.81 | — | 520 | — | — | — | $3.0\times10^{-10}$ | [85, 86] |
| $Ti_3C_2T_x$ | < 0.20 | $7.0\times10^{-1}$ | 503–652 | 1.70 | 4.10 | $-4.9\times10^{-20}$ | $-7.6\times10^{-10}$ | [87-90] |
| $CH_3NH_3PbI_3$ | 1.57–1.72 | 2.0-3.0 | 730 | 2.25 | 0 | $1.6\times10^{-12}$ | $-4.6\times10^{-7}$ | [91-95] |
| $CsPbBr_3$ | 2.32 | $6.0\times10^{-3}$ | 525 | — | — | $-1.8\times10^{-17}$ | $1.8\times10^{-11}$ | [96-99] |
| MoSSe | 1.68 | $1.6\times10^{2}$ | 695 | — | — | — | — | [100-102] |
| WSSe | 2.13 | $7.2\times10^{2}$ | 668 | — | — | — | — | [101, 103] |
| BiOBr | 2.73–2.82 | — | 450 | 2.20 | 0.20 | $3.8\times10^{-14}$ | $1.5\times10^{-7}$ | [104-106] |
| GeTe | 0.73–0.95 | $1.0\times10^{2}$ | — | 6.80 | 0.30 | — | — | [107-109] |
| Ni-MOF | 3.12 | — | 353 | — | — | $-8.9\times10^{-20}$ | $-2.5\times10^{-13}$ | [110-112] |
| Graphdiyne | 0.46–1.10 | $2.0\times10^{5}$ | 407–455 | — | — | — | $-1.6\times10^{-11}$ | [113-115] |

[a)] The carrier mobilities of all the listed materials are experimental results except for those of SnSe, $MoO_3$, MoSSe, WSSe, and Graphdiyne that are theoretical calculated values.
[b)] The *n*, *k* values for all the listed materials are at 1550 nm except for BP at 800 nm and h-BN at 1200 nm.
[c)] The NLO parameters of $PdSe_2$, $PtSe_2$, BP, h-BN, $MoO_3$, and $CsPbBr_3$ are at 800 nm. Those of $MoS_2$ and $Bi_2Te_3$ are at 1060 nm. Those of other materials are at 1550 nm.

### *2.1.1. Graphene*

Graphene is the original 2D material that has been intensively studied since 2004.[6] Its atomic structure consists of $sp^2$-hybridized carbon atoms arranged in a honeycomb lattice to form a densely packed 2D sheet.[37, 116] Hence, graphene has a gapless band structure, characterized by the so-called Dirac points where the linearly dispersive conduction and valence bands meet at the K (K') points of the Brillouin zone, resulting in a semimetal behavior.[14, 33, 37] This unique band structure yields remarkable electrical and optical properties, including extremely high carrier mobilities of ~$1.2 \times 10^5$ $cm^2$ $V^{-1}$ $s^{-1}$ under ambient conditions,[39] as well as broadband (visible to infrared) absorption that is quite high at 2.3% per layer.[38] In addition, graphene manifests attractive NLO properties, such as high saturable absorption[117] and giant optical Kerr nonlinearities.[42, 118] The gapless band structure also allows for the tuning of its chemical potential or Fermi level by changing the carrier density with a gate voltage, enabling the ultrafast dynamic tuning of its electrical and optical properties.[23, 119-121]

### *2.1.2. Graphene oxide*

Graphene oxide (GO) is a graphene derivative with a heterostructure consisting of a basal plane of carbon atoms decorated with oxygen-containing functional groups. This structure makes GO a hybrid material, featuring both conducting π-states from $sp^2$ carbon sites but at the same time with a large energy gap between the σ-states of $sp^3$-bonded carbons.[122] As the $sp^2$ fraction increases, the electrical conductivity of GO also increases, transforming from an insulator to semiconductor and graphene-like semimetal. The large bandgap of GO compared with graphene results in much lower optical absorption, particularly at infrared wavelengths, and yields broadband fluorescence in the visible and near-infrared (NIR) regions.[46, 122, 123] In addition, the bandgap of GO can readily be tuned by using various reduction methods, thus allowing for tailoring its material properties. GO also exhibits a high optical Kerr nonlinearity, several orders of magnitude larger than that of bulk materials such as silicon and chalcogenide glasses.[45, 124, 125]

### *2.1.3. Transition metal dichalcogenides*

Transition metal dichalcogenides (TMDCs), with the general formula of $MX_2$ (M represents a transition metal and X represents a chalcogen element), are a widely studied family of 2D materials. Monolayer TMDCs have a sandwich atomic structure of X-M-X, and many of them, such as $MoS_2$, $WS_2$, $MoSe_2$, and $WSe_2$, have bandgaps comparable to typical semiconductors, from 1 eV to 2.5 eV, covering a spectral range from the visible to the NIR regions.[8, 126] TMDCs manifest a carrier mobility on the order of $10^2$ $cm^2$ $V^{-1}$ $s^{-1}$.[47, 51] They also exhibit strong optical absorption (> 10% per layer) as well as efficient PL.[34, 48, 53] In addition, monolayer hexagonal TMDCs exhibit unique valley dependent properties, such as valley coherence and valley-selective circular dichroism.[8, 127] Further, TMDCs exhibit both large second and third-order optical nonlinearities that are dependent on the number of layers.[33, 128] Recently, noble metal TMDCs, such as $PdSe_2$, $PtSe_2$, and $PdTe_2$, have also attracted significant interest due to their high carrier mobility and unique anisotropic electrical and optical properties.[129, 130]

### *2.1.4. Black phosphorus*

Black phosphorus (BP) is another attractive single element 2D material that has been widely studied in the past decade. It has a puckered crystal structure, which yields a strong in-plane anisotropy in its physical properties.[10, 33, 131] BP has very high carrier mobilities that differ along armchair and zigzag crystal directions.[10, 131] Moreover, it features a thickness-dependent direct bandgap that varies from 0.3 eV (bulk) to 2.0 eV (monolayer), which covers the range between the zero-bandgap graphene and large-bandgap TMDCs.[67, 131, 132] It also exhibits an ultrahigh and anisotropic third-order optical nonlinearity that depends on layer number and light polarization.[133, 134] However, one drawback is that BP has a poor stability in ambient atmosphere, creating challenges for its use in practical applications, although from another perspective, this also provides a means to tune its material properties via localized oxidization treatments.[135]

### *2.1.5. Hexagonal boron nitride*

Hexagonal boron nitride (h-BN) has a graphene-like honeycomb lattice configuration with equal numbers of boron and nitrogen atoms. It is an insulator with a large bandgap of around 6.0 eV, corresponding to a cut-off wavelength in the ultraviolet regime.[71] Recently, single photon emission from the point defects in monolayer h-BN has been reported, which covers a spectral range from the visible to the NIR regions.[9, 136] In addition, h-BN manifests a strong NLO absorption, as well as large second and third-order optical nonlinearities.[9, 74] The excellent stability and anti-oxidation properties make the h-BN ideal for protective coatings.[71] It has also been widely used to form 2D van der Waals (vdW) heterostructures with other 2D materials such as graphene, BP, and TMDCs.[137, 138]

### *2.1.6. Other 2D materials*

Many new 2D materials have emerged in recent years, represented by MXenes, perovskites, and metal-organic frameworks (MOFs), all of which have greatly enriched the family of 2D materials.

MXenes are a family with a formula of $M_{n+1}X_nT_x$, where M is an early transition metal, X denotes C and/or N element, T represents surface terminations, and n = 1, 2, or 3.[87, 139] The electronic structures of MXenes can be modified by changing their surface functional groups,[139] allowing their bandgaps to vary from 0.05 eV to 2.87 eV, covering a broad band from the visible to the mid-infrared (MIR) regions.[140, 141] They also exhibit attractive optical properties, such as a high transmittance of visible light (> 97% per nm) as well as tunable PL wavelengths.[88, 90] Excellent NLO properties in MXenes, including broadband nonlinear absorption and a large third-order nonlinearity, have also been reported.[87, 142]

Metal-halide perovskites have a general formula of $ABX_3$, where A and B are cations with different sizes, and X represents $I^-$, $Br^-$, $Cl^-$ or mixtures.[143] Due to their prominent photovoltaic features and luminescence properties, perovskite semiconductors have been widely used for solar cells and light-emitting diodes.[144-146] In addition, perovskite materials possess a tunable bandgap from 1.2 eV to 3.0 eV through simple modification of their composition.[147] Recently, their attractive NLO properties have also been reported, such as strong saturable absorption, and a thickness-dependent Kerr nonlinearity.[13, 146, 148]

Metal-organic frameworks (MOFs) are organic–inorganic hybrid porous crystalline materials with metal ions or metal-oxo clusters coordinated with organic linkers.[11, 12] Benefiting from this unique structure, 2D MOFs exhibit many outstanding properties. For example, they have relatively large bandgaps that can be varied from 3.1 eV to 4.3 eV, allowing for many optical applications in the ultraviolet region.[149] Tunable electrical conductivities, efficient white light emission, and excellent NLO properties including broadband saturable absorption, large Kerr nonlinearities, and strong second-order nonlinearities, have also been reported for 2D MOFs.[112, 150-152]

### 2.2. Applications of integrated devices incorporating 2D materials

Motivated by the distinctive properties mentioned above, a diverse range of devices incorporating 2D materials have been demonstrated based on optical fibres,[148, 153] integrated chips,[7, 119, 154] and other platforms.[155, 156] We limit our focus on integrated devices, giving a brief overview since they have already been reviewed in detail elsewhere,[7, 14, 31, 34, 126, 157] with their applications being summarized in **Table 2.**

With atomic scale film thicknesses and high carrier mobilities, graphene, TMDCs, and BP have been widely used for field-effect transistors (FETs), especially for sub-5 nm devices,[16, 51, 158] as have other 2D materials such as GO,[159] perovskites,[91] and MXenes.[160] In addition, graphene,[161] GO,[162] $MoS_2$,[163] BP,[164] and MOFs[165] have been used to realize compact energy storage devices with high power densities, such as micro-supercapacitors and micro-batteries. Recently, highly efficient solar cells based on perovskite materials,[144, 166] and graphene/Si,[167, 168] or TMDC/Si [169] heterojunctions have been demonstrated, where 2D functional layers, such as electron/hole transport layers,[170, 171] electrodes,[172, 173] and additives,[173, 174] significantly improved their performance.

**Table 2**. Applications of integrated devices incorporating different 2D materials. FET: field-effect transistor. LED: light-emitting diode. PD: photodetector. PSD: polarization selective device. NLO: nonlinear optical.

| Material | Electronic devices | | | Optoelectronic devices | | | | Optical devices | | | Others |
|---|---|---|---|---|---|---|---|---|---|---|---|
| | FET | super-capacitor | micro-battery | solar cell | LED & laser | modulator | PD | PSD | NLO device | lens | sensor |
| Graphene | [6]<br>[16] | [161]<br>[175] | [176]<br>[177] | [167]<br>[178] | [179]<br>[180] | [23]<br>[119] | [181]<br>[182] | [26]<br>[183] | [121]<br>[184] | [185]<br>[186] | [31]<br>[187] |
| GO | [159]<br>[188] | [189]<br>[162] | — | [170]<br>[190] | [191]<br>[192] | — | [193]<br>[194] | [27]<br>[195] | [196]<br>[29] | [197]<br>[198] | [159]<br>[30] |
| $MoS_2$ | [17]<br>[199] | [163] | [200] | [201]<br>[202] | [48]<br>[203] | [204] | [154]<br>[205] | [206] | [207] | [208] | [209]<br>[210] |
| $WSe_2$ | [51]<br>[211] | — | — | [174]<br>[169] | [53]<br>[21] | — | [212]<br>[213] | — | [214]<br>[215] | [208]<br>[216] | [217]<br>[218] |
| $PdSe_2$ | [55]<br>[219] | — | — | — | — | — | [129]<br>[220] | — | — | — | [221] |
| $PtSe_2$ | [57]<br>[222] | — | — | [223] | — | [224] | [225]<br>[226] | — | — | [208] | [32]<br>[227] |
| $MoTe_2$ | [228]<br>[229] | — | — | [172] | [20]<br>[230] | [22] | [24]<br>[230] | — | — | — | [231]<br>[232] |
| BP | [66]<br>[158] | [164] | — | [171]<br>[233] | [67]<br>[234] | [235]<br>[236] | [131]<br>[132] | [237] | — | — | [238]<br>[239] |
| h-BN | — | [240] | — | [241] | [136]<br>[242] | — | [25]<br>[243] | — | — | — | [244] |
| Perovskites | [91]<br>[245] | — | — | [144]<br>[166] | [145]<br>[246] | [247] | [248]<br>[249] | — | — | [250] | [251]<br>[252] |
| MXenes | [253]<br>[160] | [254] | [255]<br>[256] | [173] | [257] | [258] | [259]<br>[260] | [261] | — | — | [262]<br>[255] |
| MOFs | [263]<br>[253] | [165]<br>[264] | — | [265] | — | — | [266] | — | — | — | [267]<br>[263] |
| Graphdiyne | [268] | [269]<br>[270] | — | [269]<br>[270] | [115] | — | [271] | — | — | — | [272] |

By utilizing the extraordinary luminescent properties of perovskites,[145] GO,[191] TMDCs,[203, 230] and BP,[234] light-emitting diodes (LEDs) with broadband emission from the visible to the MIR regions have been realized. Efficient laser emission from nano-resonant cavities integrated with monolayer $WSe_2$ and $MoTe_2$ has been achieved.[20, 21] With strong light absorption and low dispersion, graphene, reduced graphene oxide (rGO), BP and TMDCs have enabled broadband photon detection from the ultraviolet to the terahertz regions.[25, 129, 131, 182, 194] Integrated optical modulators have also been demonstrated, including electro-optic,[23, 204, 235] thermo-optic,[224, 273] and all-optical modulators.[22, 274] By virtue of their low dispersion and large anisotropic light absorption, broadband polarization-selective devices based on graphene,[26] GO,[27] $MoS_2$,[206] BP,[237] and MXenes[261] have been realized. 2D materials with superior NLO properties have been used for implementing high-performance all-optical signal processing devices based on four wave mixing (FWM),[184, 196] self-phase modulation,[29, 207] and second-harmonic generation.[214] Recently, gate-tunable frequency comb generation was also demonstrated based on a graphene/silicon nitride hybrid microring resonator (MRR).[121] The patterning of ultrathin graphene,[186] GO,[198] TMDCs,[208, 216] and perovskite[250] films has formed the basis of flat optical lenses. Chip-scale sensors incorporating 2D materials have also demonstrated the detection for a wide range of targets, from gases,[30, 231] humidity,[227, 272] and metal ions,[31, 209] to pressure,[32, 262] and biomolecules.[31, 263]

## 3. Integration fabrication techniques

As discussed in Section 2, atomically thin layers manifest extraordinary properties compared to bulk crystals, thus enabling many new hybrid integrated devices. In this section, we summarize and discuss the fabrication techniques for integrating 2D materials. **Fig. 2** shows the outline of this section, which includes four subsections – material synthesis, on-chip transfer, film patterning, and property tuning / modification.

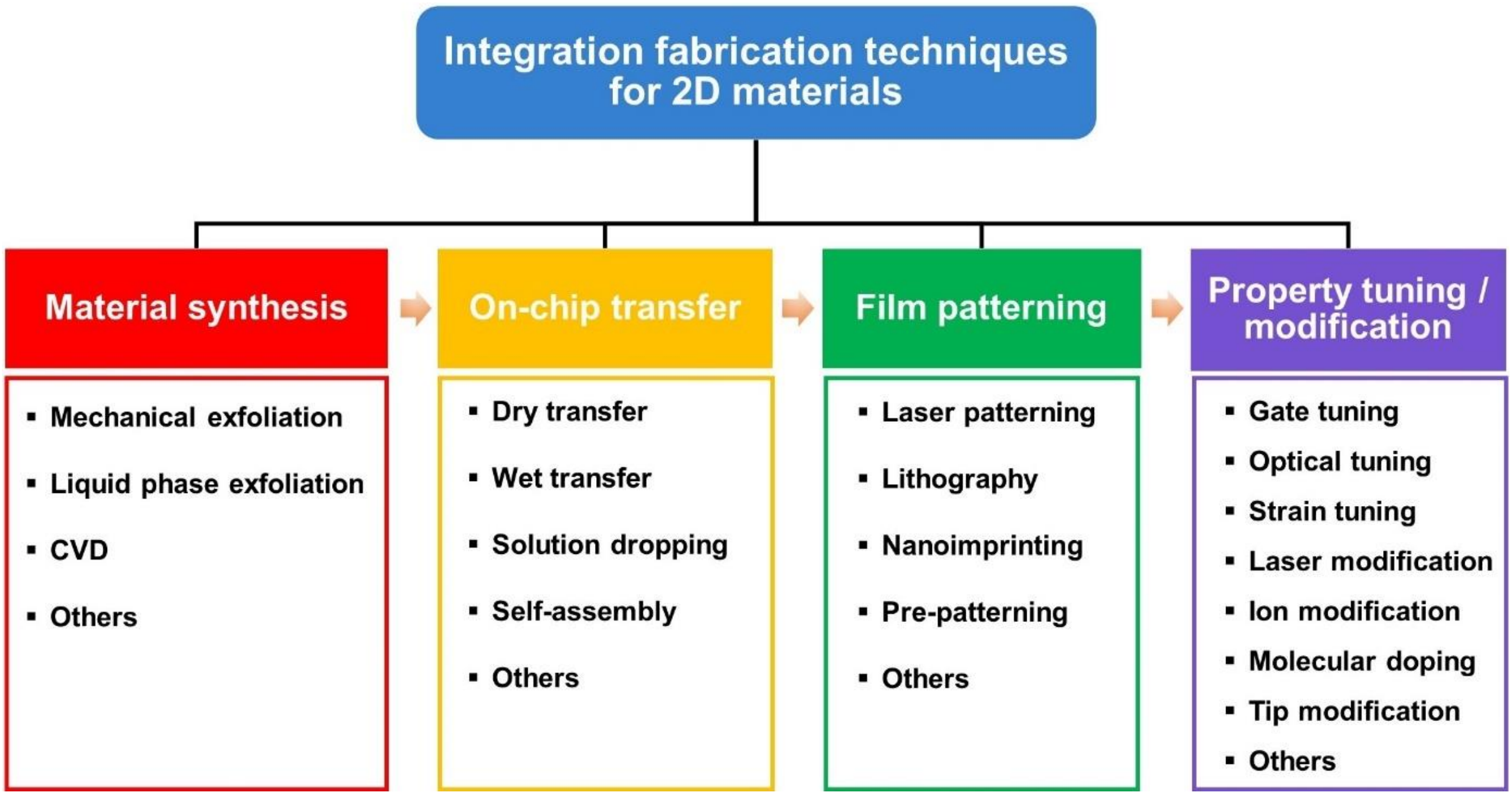

**Figure 2.** Outline for Section 3 on integration fabrication techniques for 2D materials. CVD: chemical vapor deposition.

### 3.1. Material synthesis

Sythesizing high quality 2D materials is the first step for their on-chip integration. Here, we sumarize the synthesis methods and discuss their potential for scalable and industrial applications.

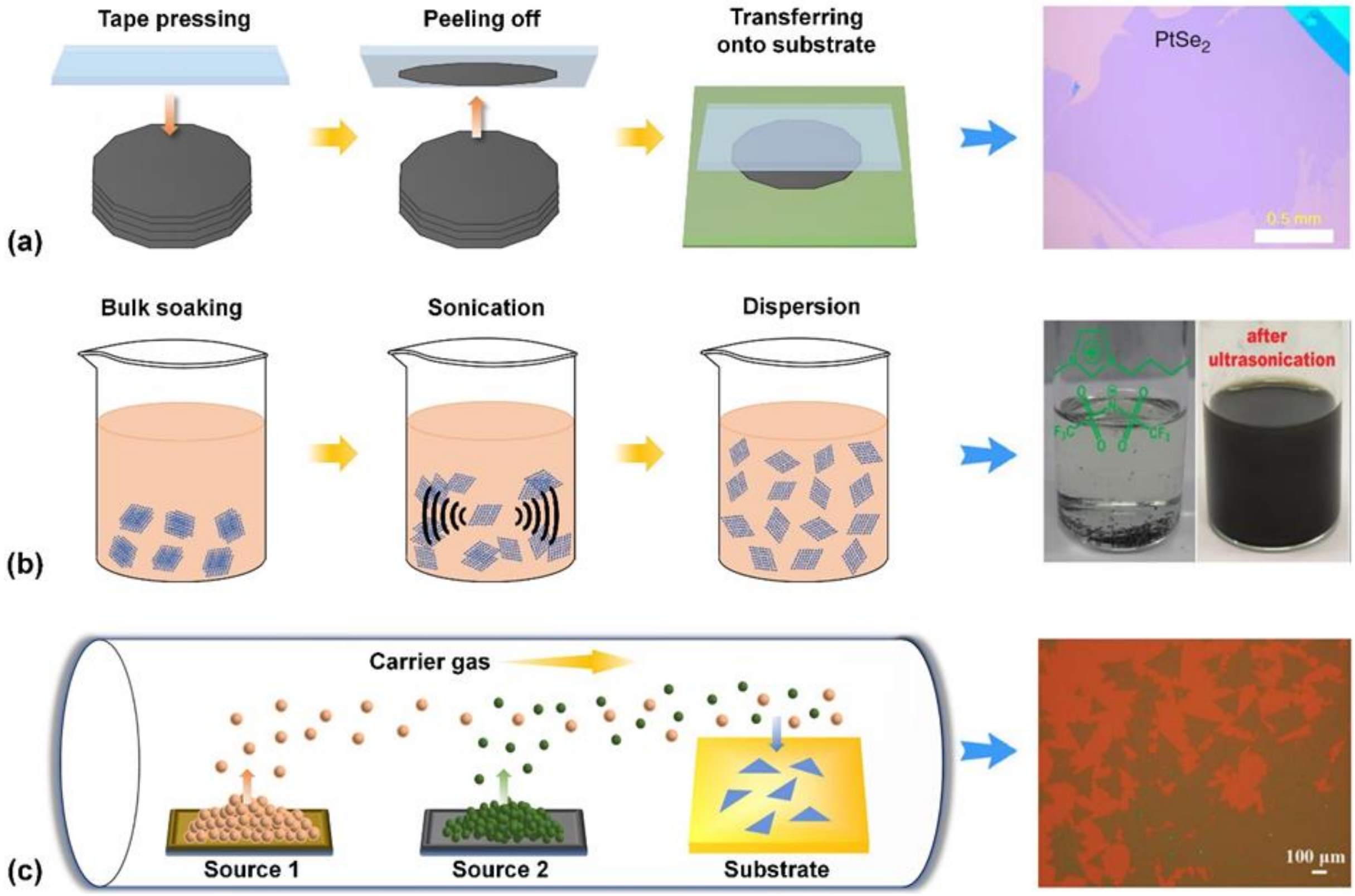

**Figure 3.** Typical synthesis methods of 2D materials. (a) Schematic illustration of mechanical exfoliation (ME) processes. Right: a microscope image of a mechanically exfoliated $PtSe_2$ flake on $SiO_2$/Si substrate. Reproduced with permission.[275] Copyright 2020, Springer Nature. (b) Schematic illustration of liquid phase exfoliation (LPE) processes. Right: images of graphite in ionic liquids before (left) and after (right) ultrasonication. Reproduced with permission.[276] Copyright 2010, Royal Society of Chemistry. (c) Schematic illustration of chemical vapor deposition (CVD) processes. Right: a microscope image of CVD-grown $MoS_2$ crystals on $SiO_2$/Si substrate. Reproduced with permission.[17] Copyright 2016, Wiley-VCH.

### *3.1.1. Mechanical exfoliation*

Monolayer graphene[6] was first obtained by mechanical exfoliation (ME) using adhesive tape, as shown in **Fig. 3(a)**. By repeated peeling off a small piece of bulk, single crystal graphene, monolayer or few-layer samples can be straightforwardly obtained. However, the random thicknesses and small flake sizes yielded by this process restrict its applications to the laboratory rather than production scale uses. Hence, there has been significant motivation to modify the conventional ME method. One alternative approach has been to introduce a polymeric intermediate substrate to improve the exfoliation yield and lateral size of the exfoliated 2D flakes.[277, 278] Gold-assisted ME methods have also been used to improve the exfoliation yield. Due to the strong interaction between Au and the surface terminating elements of most 2D materials, it is easy to exfoliate large-area (up to the centimeter scale) monolayers or few-layer flakes onto an Au adhesion substrate.[199, 275] After post-etching the Au film, the resulting 2D flakes can then be transferred to integrated chips. A more general method based on a gold tape was recently reported,[279] where the tape contained a top thermal release tape, a polyvinylpyrrolidone (PVP) interfacial layer, and an ultra-flat Au bottom layer. This method can disassemble a wide range of 2D crystals layer by layer into single-crystal monolayers, with lateral dimensions limited only by the bulk size. Additionally, by using Ni as an assisting layer, a layer-resolved splitting method was employed to prepare wafer-scale monolayer h-BN and TMDCs.[280]

#### *3.1.2. Liquid phase exfoliation*

Liquid phase exfoliation (LPE) is another widely used technique. In contrast to ME, LPE is a solution-based method that can yield a large volume of 2D nanosheets in a cost-effective way. However, some drawbacks, such as relatively poor sample qualities and small flake sizes, still limit its applications. **Fig. 3(b)** shows a schematic illustration of typical LPE processes, where bulk crystals are first soaked by a specific solvent. Sonification or shear mixing is then utilized to provide a mechanical force that exfoliates the bulk into dispersed 2D nanosheets.[281-283] The lateral size and thickness of the exfoliated flakes can be controlled by optimizing the sonication and mixing parameters. In LPE processes, obtaining a suitable surface tension of the solvent is critical to decrease the exfoliation energy and stabilize the exfoliated nanosheets.[284] By using this method, monolayer and few-layer nanosheets of a wide range of 2D materials, such as graphene, h-BN, and TMDCs, have been fabricated.[281, 282] To weaken interlayer binding and reduce the exfoliation energy barrier, ion intercalation has also been introduced to enable scalable exfoliation of layered crystals.[283-285] For example, ionic compounds have been used to intercalate the graphite to produce a bilayer or trilayer graphene solution.[285] Li ion intercalation has also been employed to produce high-yield liquid exfoliation of 2D nanosheets such as TMDCs and some metal oxides.[283, 284, 286] Recently, the intercalation technique was also employed to induce doping ions to 2D hosts during the LPE process,[287, 288] thus providing a simple approach with a large yield to engineer the electronic properties of these crystal lattices. For materials that contain an exchangeable interlayer of cationic counterions, such as $TiO_2$, an ion exchange strategy was developed to exfoliate their bulk crystals.[283] In addition, modified Hummer's methods, termed oxidation-assisted LPE,[289] and chemical-etching-assisted LPE,[290] have been developed to fabricate GO and MXene 2D nanosheets, respectively.

### *3.1.3. Chemical vapor deposition*

The direct growth of 2D materials is another synthesis method that is attractive for industrial scale production. Chemical vapor deposition (CVD) is a powerful approach to synthesize 2D films on a range of substrates, and this has been used to fabricate large-area 2D films with controllable thicknesses.[17, 291, 292] In CVD processes (**Fig. 3(c)**), a tube furnace is used, in which specific precursors are transported by the carrier gas, after which they react together to form a 2D film on a flat substrate, such as Au or sapphire. To control the film size and thickness, the mass flux and growth rates need to be accurately controlled. The nucleus formation and growth of the 2D crystal domains are determined by the mass flux, while the grain size of the as-grown films is set by the growth rate.[291] Recently, molten-salt-assisted technology[291] and gold alloyed metal film solid solution technology[292] have been used in synthesizing 2D TMDCs. By accurately controlling the deposition temperature, fold-free large-area single crystal graphene films have been fabricated on Cu-Ni alloyed foils.[293] Efforts to directly grow 2D films onto integrated substrates, such as silicon and silica,[291, 294] have further advanced CVD techniques. In addition, by evaporating different dopants during the CVD growth process, in-situ substitutional doping of both cations and anions has also been achieved.[288, 295]

### *3.1.4. Other synthesis methods*

Molecular beam epitaxy (MBE) has been used to fabricate high quality, large-area (wafer scale) monolayer 2D materials, including graphene, TMDCs, and h-BN.[72, 296, 297] This epitaxial growth method allows for in-situ characterization of the synthesized films. Physical vapor deposition (PVD) is another popular technique that, in contrast to CVD and MBE, provides higher deposition rates useful for thicker or multilayer films.[59, 298, 299] Template method is another strategy to fabricate 2D nanosheets and their aqueous dispersions of MXenes.[300] Solution based synthesis methods that involve different organic or inorganic precursors have been widely used to deposit perovskite thin films.[97, 301] Other approaches, such as interfacial synthesis methods (ISM) and surfactant-assisted methods (SAM), are used in the fabrication of 2D MOF nanosheets.[35, 302]

**Table 3.** Synthesis methods of different 2D materials and corresponding sample features [a)].

| Material | Synthesis method [b)] | Sample form / feature size |
|---|---|---|
| Graphene | ME,[37] LPE,[282] CVD,[293, 303] | 1. Single crystal monolayer flakes / millimeter scale [305] |
| | MBE,[296, 304] MAME [305] | 2. Deposited single crystal monolayer films / wafer-scale [c)] [296, 303] |
| | | 3. Nanosheets in solution / micrometer scale [282] |
| GO | MHM,[289, 306] ECM,[307] HTM,[308] | 1. Nanosheets in solution / micrometer scale [289] |
| | LPE,[309] CVD [310] | 2. Deposited multi-layer films / millimeter scale [310] |
| TMDCs | ME,[47] LPE,[284, 311] CVD,[291] | 1. Single crystal monolayer flakes / centimeter scale [275] |
| | MOCVD,[312] MBE,[297] PVD,[299] | 2. Deposited monolayer films / wafer-scale [312] |
| | MAME [275, 279] | 3. Nanosheets in solution / micrometer scale [311] |
| BP | ME,[158] LPE,[313, 314] MAME,[275] | 1. Single crystal monolayer flakes / millimeter scale [275] |
| | PTM [315, 316] | 2. Deposited multi-layer films / millimeter scale [315] |
| | | 3. Nanosheets in solution / micrometer scale [314] |
| h-BN | ME,[137] LPE,[317] PVD,[298] | 1. Single crystal monolayer flakes / micrometer scale [320] |
| | CVD,[318, 319] MBE,[72] MAME [280] | 2. Deposited single crystal monolayer films / wafer-scale [319] |
| | | 3. Nanosheets in solution / micrometer scale [317] |
| MXenes | CELPE,[290, 321] TM,[300] CVD,[322] | 1. Single crystal multi-layer flakes / micrometer scale [322] |
| | PVD [323] | 2. Deposited multi-layer films / — [323] |
| | | 3. Nanosheets in solution / micrometer scale [290] |
| Perovskites | ME,[324] LPE,[325] SSM,[97, 301] | 1. Single crystal multi-layer flakes / micrometer scale [324] |
| | CVD,[148, 166] PVD [326] | 2. Deposited multi-layer films / centimeter scale [166] |
| | | 3. Nanosheets in solution / micrometer scale [325] |
| 2D-MOFs | ME,[327] LPE,[328, 329] ISM,[302] | 1. Single crystal multi-layer flakes / micrometer scale [327] |
| | SAM,[35, 330] CVD [331] | 2. Deposited multi-layer films / centimeter scale [331] |
| | | 3. Nanosheets in solution / micrometer scale [35, 329] |

[a)] Here, we only summarize the main synthesis methods and typical sample features.
[b)] ME: mechanical exfoliation. LPE: liquid phase exfoliation. CVD: chemical vapor deposition. MBE: molecular beam epitaxy. MAME: metal-assisted mechanical exfoliation. MHM: modified Hummer’s method. ECM: electrochemical method. HTM: hydrothermal method. MOCVD: metal-organic chemical vapor deposition. PVD: physical vapor deposition. PTM: phase transformation method. CELPE: chemical-etching-assisted liquid phase exfoliation. TM: template method. SSM: solution synthesis method. ISM: interfacial synthesis method. SAM: surfactant-assisted method.
[c)] Here wafer-scale means the film lateral size is larger than 2 inches.

**Table 3** summarizes the main synthesis techniques as well as their typical sample forms and feature sizes. In general, each 2D material has a unique range of synthesis methods to fabricate samples varying from single crystal flakes to dispersed nanosheets and large-area films. Techniques such as ME, LPE, and CVD, are applicable to a wide range of materials while other approaches are more specialized. 2D nanosheets in solutions typically have micrometer lateral sizes while the feature size of CVD grown films can be as large as several tens of centimeters. For graphene, TMDCs, BP, and h-BN, large-area (up to the wafer scale) single crystal monolayers can be fabricated either via metal-assisted ME (MAME) or MBE or CVD. On the other hand, for some of the newly developed materials such as MXenes and perovskites, the sample size is limited to several micrometers even with state-of-the-art synthesis methods, mainly because of their complicated material phase diagrams.[332]

Synthesizing 2D films on wafer scales, especially in single crystal monolayer form, is important for the large-scale fabrication of integrated devices, for which MAME, MBE, and CVD have attracted significant attention. Fabrication efficiency is particularly important for industrial scale applications, and in this respect the low deposition speed of MBE represents a significant limitation, as does the manual manipulation nature of current MAME techniques. In contrast, CVD, widely used in the semiconductor industry, has a strong potential for cost-effective production for graphene and TMDCs, although the direct growth of high-quality 2D films on integrated wafers is still under development. In addition, LPE is a facile fabrication process with a high yield efficiency, and is promising for the mass production of 2D nanosheets such as GO and BP.

### 3.2. On-chip transfer techniques

Transferring synthesized 2D films onto desired substrates is the second step for the on-chip integration of 2D materials. Different transfer techniques have been developed for various synthesis methods, including dry transfer, wet transfer, solution dropping, self-assembly, and others, each with different capability and controllability serving for varied application demands.

#### *3.2.1. Dry transfer*

**Fig. 4(a)** shows a schematic illustration of dry transfer processes that have been developed to stack 2D flakes onto a desired substrate. A 2D flake is first exfoliated onto a polydimethylsiloxane (PDMS) transfer stamp, which is then attached to the target surface with a micromanipulator, with an optical microscope used to identify the targeted flake and control its position. The flake is finally detached from the stamp and adhered to the target position after very slowly peeling off the stamp. To achieve a successful transfer, the attaching pressure of the stamp onto the target surface as well as the peeling off speed need to be accurately controlled.[15, 333] For materials with poor stability in atmosphere, such as BP, the transfer processes can be performed in a vacuum glove box. Recently, a so-called pick-up technique[334-336] was developed to directly pick up a single flake and transfer it onto a target position. Special stamps were designed for the transfer processes, such as a polypropylene carbonate (PPC) coated PDMS block[334] and hemisphere PDMS stamps.[335, 336] This technique enables multi-dimensional manipulation of the flakes, such as rotation controls or twisted stackings, and alignments along specific crystal directions.[334-336]

The dry transfer technique is applicable to most layered materials and does not involve chemical liquids, therefore avoiding chemical pollution from the residuals and ensuring a better interaction between the 2D layers and target substrates.[24, 336] On the other hand, it is limited in terms of the lateral size of the exfoliated flakes it can process, and so is not suitable for large-area film transference, particularly for conformal coatings on substrates with complex structures such as waveguides or gratings. A relatively low success rate and the requirements for complex support facilities also limit its transfer efficiency and hence production yield.

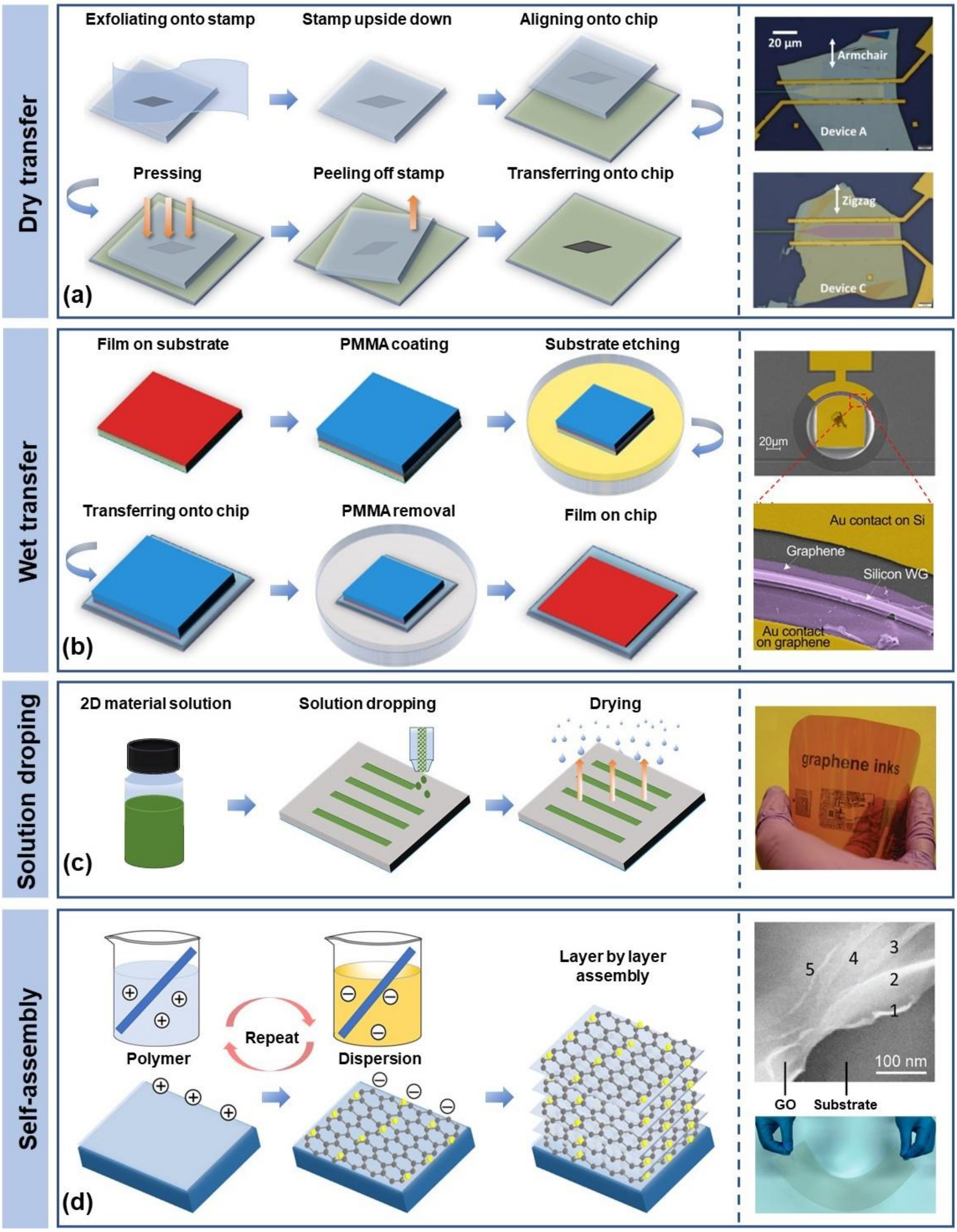

**Figure 4.** Typical on-chip transfer techniques of 2D materials. (a) Schematic illustration of dry transfer processes. Right: microscope images of as-transferred BP flakes on a silicon grating coupler. Reproduced with permission.[337] Copyright 2018, American Chemical Society. (b) Schematic illustration of wet transfer processes. Right: microscope images of a silicon microring resonator (MRR) covered by an as-transferred graphene film. Reproduced with permission.[338] Copyright 2015, American Chemical Society. (c) Schematic illustration of solution dropping processes. Right: an image showing large-area inkjet-printed graphene patterns on polyimide substrate. Reproduced with permission.[339] Copyright 2015, Wiley-VCH. (d) Schematic illustration of in-situ self-assembly processes. Right: a scanning electron microscope (SEM) image of a five-layer GO structure on a silicon substrate (up) and a large-area coating of GO film on a flexible transparent substrate (down). Reproduced with permission.[44] Copyright 2019, American Chemical Society.

### *3.2.2. Wet transfer*

As discussed above, CVD is capable of synthesizing large-area 2D films, and so is attractive for industrial scale applications. However, the as-synthesised films are usually grown on metal substrates or foils rather than dielectric materials such as silicon or silica that are used for integrated devices. Hence, subsequent film transfer processes are often needed such as wet transfer techniques. **Fig. 4(b)** shows a schematic of these processes. In the first step, a poly (methyl methacrylate) (PMMA) layer is spin coated onto the 2D film grown on a substrate. Next, the PMMA/film/substrate stack is placed into a chemical solution, such as $FeCl_3$ for graphene transfer from Cu substrate[23, 126] and KOH for TMDC film transfer from $SiO_2$/Si or sapphire substrates,[340, 341] to remove the substrate. After removing the substrate, the PMMA/film stack is cleaned and attached to a target substrate. The whole structure is then soaked in acetone solution to dissolve the PMMA layer. Finally, after washing and drying, the 2D film is successfully transferred. In addition to substrate etching, other strategies have been used to detach CVD-grown 2D films from their deposition substrates. For example, the bubbling method based on water electrolysis procedures has been used to delaminate monolayer graphene and $WS_2$ from the Au and Pt substrates, respectively.[342, 343] A surface-energy-assisted strategy has also been developed to remove the substrates from $MoS_2$ films, in which a polystyrene (PS) layer acts as a transfer medium and the water penetration between the $MoS_2$ films and the substrates provides the detachment forces.[344, 345]

Compared with dry transfer methods that only support the transfer of small flakes, wet transfer techniques can operate on large-area films. Moreover, they are easy to operate and do not rely on sophisticated facilities, therefore yielding a high success rate and transfer efficiency, making it appealing for potential industrial applications. However, this method does have drawbacks. The chemical solutions and PMMA residuals can contaminate the substrate surface and weaken the interaction between the 2D layers and functional structures, thus degrading device performance.[126] In addition, due to stretching, bending, and wrinkling that may occur during the transfer process, it is difficult to achieve a high film uniformity, or conformal coating of substrates featuring complex structures. The film imperfectness and the lack of assisted facilities for precise position identification also make it challenging to fabricate multilayered films and control crystal orientation.

### *3.2.3. Solution dropping*

2D nanosheets prepared by LPE can be directly transferred via solution dropping methods, such as drop casting,[346, 347] spin and spray coating.[163, 284] These methods provide a rapid and simple approach to prepare 2D films onto integrated chips. Recently, inkjet printing has also been introduced due to its low cost, additive patterning, and scalability for large-area manufacturing.[254, 314] **Fig. 4(c)** shows a schematic of the solution dropping processes, where three main steps are involved, including high-quality ink solution preparation, programmed ink jetting, and drying treatment. Successful printing involves a number of key aspects, including the formation of a stable jet of single droplets, the ability to adequately wet the substrate, and realization of a uniform distribution of material during the droplet drying process.[314] Different solvents are used for the various kinds of inks. For example, Li *et al.* introduced a distillation-assisted solvent exchange strategy to prepare dispersed graphene in high viscosity terpineol and achieved a well-directed and constant jet stream of graphene.[348] A general approach based on isopropanol (IPA) and alcohol has been demonstrated that has achieved uniform printing of 2D crystals, including graphene, TMDCs, and BP, and that has enabled scalable and even wafer-scale device fabrication.[349] In these approaches, rapid and uniform droplet drying is critical, and additives such as polymer binders are often included into 2D crystal solutions to improve the uniformity of the printed films during evaporation. High temperature annealing, or intense pulsed light irradiance, has been employed to assist in removing these binders.[314, 339]

Solution dropping approaches provide a powerful, convenient, and robust method to transfer 2D materials that is compatible with LPE and suitable for large-scale fabrication.[254]

Assisted by programmed jetting, maskless patterning of 2D films can readily be achieved, although this method also has its own drawbacks. Its biggest limitation arises from the relatively low film uniformity and large film thicknesses that it can achieve, limiting its range in capability for fabricating high precision devices.[19, 349] To improve the film uniformity, repeated solution dropping is often performed. Typical film thicknesses are on the order of hundreds of nanometers, and it is very challenging to fabricate very thin films (< 100 nm) with this approach.[348]

***3.2.4. Self-assembly***

Self-assembly is another solution-based 2D material transfer technique, which can be realized based on a range of mechanisms, such as electrostatic attachment, hydrogen binding, covalent bonding, and ionic charge transfer.[44, 350] **Fig. 4(d)** shows an example of self-assembly based on electrostatic attachment. Here, the layer-by-layer transfer process can be divided into several steps. First, a dispersion composed of negatively charged 2D nanosheets is prepared. These nanosheets will attach to any positively charged surface through electrostatic forces while the opposing electrostatic forces between them prevent their aggregation. Next, a target substrate is immersed in an aqueous polymer, for example a polyelectrolyte polydiallyldimethylammonium chloride (PDDA) solution, to obtain a positively charged surface. The PDDA-coated substrate is then put into the prepared 2D nanosheet solution with optimized concentration and immersion time. Finally, a monolayer of the 2D material is formed onto the target substrate. By repeating these steps, multilayer films with a precisely controlled number of layers can be fabricated. To date, self-assembly techniques have been successfully employed for a range of materials, such as graphene, GO, TMDCs, and MXene films.[27, 29, 351-353]

Self-assembly is highly scalable since the area of the film is only limited by the solution container size.[44] The layer-by-layer transfer process also allows the precise control of the layer numbers, thus enabling film thickness control on the nanoscale as well as conformal coating of complex device structures. Currently, self-assembly techniques are mainly limited to laboratory-based applications – their industrial scale implementation is still at an early stage.

One major limitation of this approach is its low overall efficiency – even for very thin films, solutions with much larger volumes of 2D nanosheets are required to guarantee uniform coating.[350] Other drawbacks include its time consuming nature as well as limitations in the overall film thickness it can achieve, which make it challenging to fabricate thick (> 1 μm) films. However, these are not fundamental limitations and could be mitigated by industry automation and reuse of the GO solution.

### *3.2.5. Other transfer techniques*

As discussed above, wet transfer methods can operate on large-scale CVD-grown films but do not allow the precise positioning of single crystal flakes. Dry transfer processes, on the other hand, can achieve this. In these approaches, flakes are attached to a PDMS stamp either by direct exfoliation or a pick-up procedure, which is not applicable to CVD-grown flakes because of the strong adhesive forces between the flakes and substrate. Recently, a hybrid technique – a so-called semi-dry transfer approach – has been developed to precisely transfer CVD-grown 2D flakes in two steps. The first step is similar to a wet transfer process where CVD-grown flakes are detached from the substrate and transferred to a PDMS stamp, assisted by sacrificial polymers such as PMMA and polyvinyl alcohol (PVA).[354, 355] Next, after washing and drying, the prepared stamp is used to transfer the CVD flakes onto a target substrate in a dry transfer process. This approach is attractive for its ability to precisely transfer CVD-grown flakes although it does have drawbacks that are common to both wet and dry transfer processes, such as the limited flake sizes it can yield, the possibility of chemical solution contamination, the sophisticated nature of the transfer steps, and the low overall efficiency.

In contrast to transferring as-synthesized 2D flakes or films onto substrates, the ability to directly grow 2D layers onto designed structures is very attractive. It avoids any chemical pollutants or material deformation that often occurs with solution printing or transfer processes.[356] Recently, atmospheric pressure CVD was used to directly grow graphene on silica.[357] A modified two-step CVD method has also been employed to grow monolayer $MoS_2$ on silica structures.[358] In addition, by using a thermally assisted conversion and atomic layer deposition (ALD) method, area-selective growth and conformal coating of $PtSe_2$ films onto substrates having different topologies, including trenched substrates and striped waveguides, have also been demonstrated.[359]

**Table 4.** Comparison of on-chip transfer techniques of 2D materials. CVD: chemical vapor deposition. PDMS: polydimethylsiloxane. PMMA: poly (methyl methacrylate).

| Technique | Sample | Medium [a)] | Coating area | Conformal coating | Uniformity | Yield | Ref. |
|---|---|---|---|---|---|---|---|
| Dry transfer | Exfoliated flakes | PDMS | Small | No | High | Low | [333, 360] |
| Semi-dry transfer | CVD-grown flakes | PDMS/sacrificial polymer | Moderate | No | Moderate | Low | [354, 355, 360] |
| Wet transfer | CVD-grown films | PMMA | Large | No | Low | High | [126, 340] |
| Solution dropping | Nanosheet dispersion | — | Large | No | Low | High | [19, 254, 314] |
| Self-assembly | Nanosheet dispersion | — | Large | Yes | Moderate | Moderate | [44, 351] |
| Direct growth | — | — | Large | Yes | High | Moderate | [359, 361] |

[a)] Here, the medium is the main auxiliary polymer used in the transfer process.
[b)] Conformal coating determines the capability of transferring 2D films onto target substrates with unflatten surfaces or complex structures.

A comparison of different transfer techniques is shown in **Table 4**. Auxiliary media, such as PDMS and PMMA, are needed in both dry and wet transfer processes. Wet and semi-dry transfer approaches are mainly used for transferring CVD-grown 2D films. For 2D nanosheets synthesized via LPE, both solution dropping and self-assembly can be used for directly coating them onto integrated chips without using auxiliary media. For practical device fabrication, the combined abilities to achieve large-area coating, conformal coating over complex device structures, and high film uniformity are all highly desirable. Most techniques in **Table 4** can achieve large-area coating, except for dry transfer approaches where the area is mainly limited by the typically small lateral size of the exfoliated 2D flakes. For conformal coating, approaches such as self-assembly and direct growth that do not involve complex layer transfer processes are preferable over the others. The film uniformity achieved with wet transfer and solution dropping methods is the poorest of all methods due to imperfections such as film wrinkling, voids, and layer overlap introduced during the transfer process.

## 3.3. Film patterning

The ability to pattern 2D materials is critical for fabricating advanced integrated devices. Here, we discuss the methods used to achieve this, including laser patterning, lithography, nanoimprinting, pre-patterning, and other emerging techniques.

### *3.3.1. Laser patterning*

Laser patterning has been widely used to fabricate 2D and 3D structures in polymers, metal surfaces, and bulk materials for several decades,[362-365] and has recently been applied to 2D materials. There are two main laser-based patterning methods – direct laser writing (DLW)[366, 367] and laser interference patterning.[368, 369] DLW provides a single-step, mask-free, and chemical-free method to realize controlled thinning and patterning of 2D materials.[370-372] **Fig. 5(a)** shows a schematic illustration of DLW processes. Either continuous wave (CW) or pulsed lasers can be employed as the laser sources. Writing with CW lasers mainly involves photothermal accumulation, while pulsed lasers with less thermal effect have much higher peak intensities and so can involve more complex light-matter interaction, such as NLO absorption.[373, 374] The minimum feature size of the fabricated patterns is mainly limited by the spot size of the focused laser beam.[371] Different 2D materials involve their own unique physical mechanisms for laser patterning. Laser thinning or ablation has been widely used to pattern 2D films such as graphene, GO, TMDCs, perovskites, and MXenes, where active film regions are thinned or completely removed either by localized thermal melting or laser-induced lattice sublimation or a combination of these effects.[370, 375, 376] Photochemical reactions, such as laser-induced reduction and oxidation, have also been employed to pattern GO, graphene, and BP.[135, 371] Its simple fabrication process, together with the capability of patterning a wide range of materials, makes DLW attractive for scalable industrial applications. Some pioneering work has already emerged, such as logic circuit fabrication[18] and sensor array patterning.[377] Laser interference has also been used for 2D film patterning, which involves the interference between laser beams that results in a periodic intensity distribution, thus allowing for fast fabrication of periodic patterns over large areas, such as gratings and flower-like arrays.[368, 378]

**Figs. 5(b)** – **(m)** show film patterns for different 2D materials fabricated via laser patterning. Monolayer graphene nanoribbons (**Fig. 5(b)**) were formed via femtosecond laser ablation,[366] achieving a resolution width down to 400 nm by optimizing the laser fluence and translation speed. Graphene nanodisk arrays (**Fig. 5(c)**) and $MoS_2$ nanohole arrays (**Fig. 5(d)**) were fabricated via CW laser-based DLW, in which an Au nanoparticle layer was introduced as a thermo-plasmonic substrate to reduce the required laser power.[379] In contrast, **Figs. 5(f)** and **(g)** show curvilinear and comb-like microcircuits fabricated by laser reduction of GO films using femtosecond pulsed lasers, achieving a feature size down to 500 nm, with tunable electrical resistivities achieved by adjusting the laser power.[363] This same method has been used to fabricate ultrathin GO flat lens (**Figs. 5(h)** and **(i)**).[197, 380]

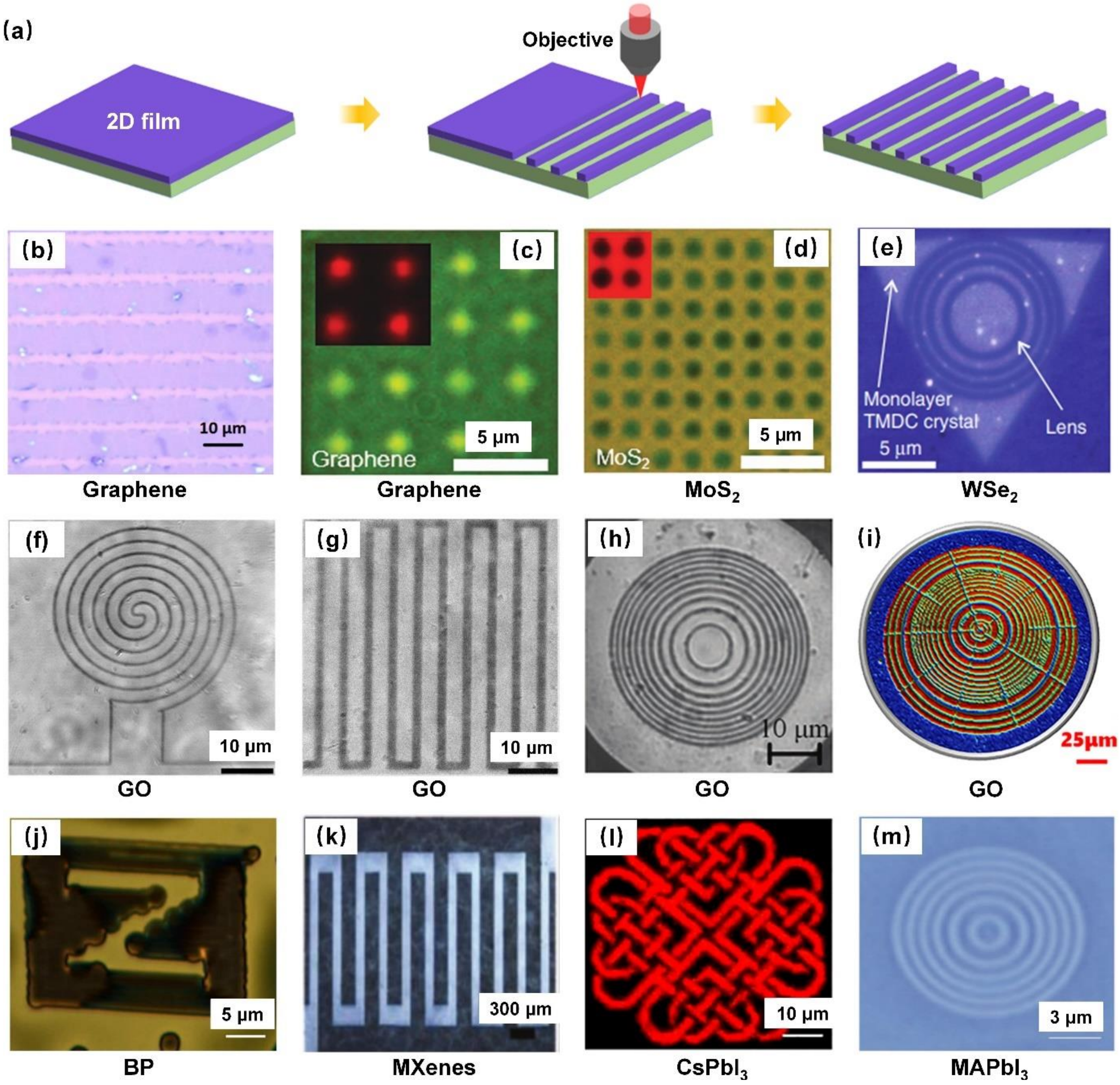

**Figure 5.** Laser patterning of 2D materials. (a) Schematic illustration of direct laser writing (DLW) processes. Optical images of laser patterned (b) Graphene nanoribbons. Reproduced with permission.[366] Copyright 2014, AIP Publishing. (c) Graphene nanodisk array. (d) $MoS_2$ nanohole array. Reproduced with permission.[379] Copyright 2018, Wiley-VCH. (e) Monolayer $WSe_2$ lens. Reproduced with permission.[208] Copyright 2020, Springer Nature. (f) and (g) GO microcircuits. Reproduced with permission.[363] Copyright 2010, Elsevier Ltd. (h) rGO flat lens. Reproduced with permission.[380] Copyright 2019, American Chemical Society. (i) A stretchable GO metalens. Reproduced with permission.[197] Copyright 2021, American Chemical Society. (j) A micropattern on a BP flake. Reproduced with permission.[135] Copyright 2015, American Chemical Society. (k) MXene microcircuit. Reproduced with permission.[381] Copyright 2020, Wiley-VCH. (l) Chinese knot pattern of $CsPbI_3$ under UV-365 nm light illumination. Reproduced with permission.[382] Copyright 2021, American Chemical Society. (m) 2D $MAPbI_3$ flat lens. Reproduced with permission.[250] Copyright 2020, Wiley-VCH.

Film patterning of BP based on laser oxidation has been demonstrated, as shown in **Fig. 5(j)**.[135] The BP oxide patterns exhibited striking optical properties where different colours were emitted when excited by light at different wavelengths. Recently, DLW has been used in the fabrication of MXene and 2D perovskite patterns.[381-383] **Fig. 5(k)** shows a microcircuit fabricated from MXenes, as a supercapacitor.[381] In-situ patterning of $\gamma$-$CsPbI_3$ quantum dot based films (**Fig. 5(l)**) was also achieved,[382] which exhibited bright red PL with a high quantum yield of up to 92%. Finally, ultrathin flat lenses in 2D perovskite (**Fig. 5(m)**) were fabricated by using a femtosecond laser,[250] where the feature size of the lens rings was ~400 nm.

### *3.3.2. Lithography techniques*

The mainstream integrated circuit industry employs well-developed lithography techniques for device patterning that include photolithography and electron beam lithography (EBL).[384] These techniques have been applied to pattern 2D materials[385-387] and offered promising routes to accelerate the industrial scale fabrication of integrated devices incorporating them. Here, we distinguish these methods according to whether they are based on: (1) lithography and etching or (2) lithography followed by lift-off.

**Fig. 6(a)** illustrates a typical process flow for the lithography & etching processes. A resist layer is first spin-coated onto a target 2D film, followed by patterning the resist via lithography. The patterning resolution is generally higher for EBL than for photolithography, although it requires longer exposure times. Therefore, photolithography, including ultraviolet, deep ultraviolet (DUV), and extreme ultraviolet (EUV) lithography, is more widely used for industrial scale, high volume throughput fabrication. In photolithography the patterned resist acts as a mask for the 2D film for subsequent etching processes, where typical dry etching methods such as reactive ion etching (RIE) or reactive gas etching are employed to etch the uncovered 2D films, thus allowing the pattern in the resist to be transferred to 2D films.

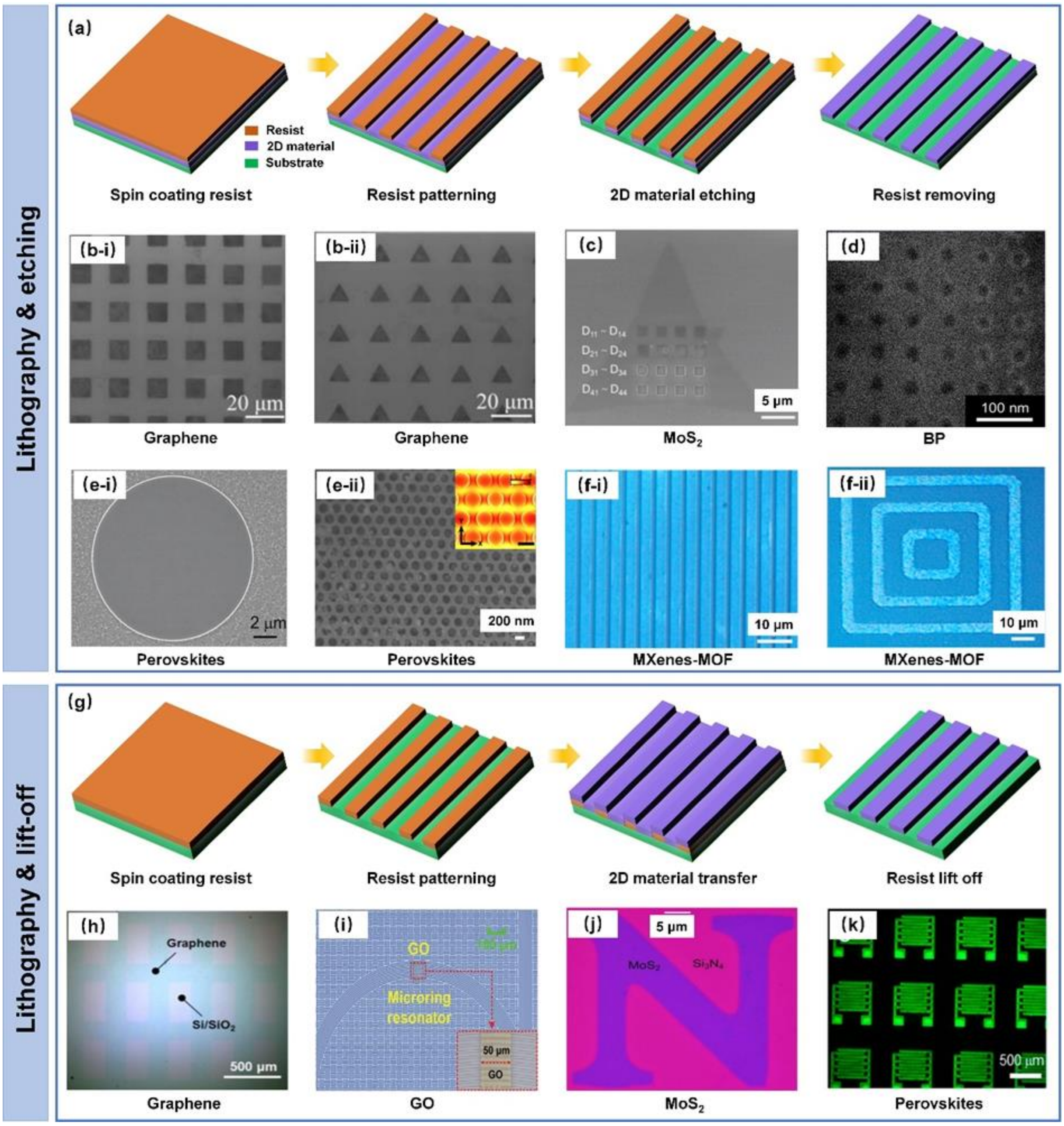

**Figure 6.** Lithography patterning of 2D materials. (a) Schematic illustration of patterning processes via lithography & etching. (b) SEM images of fabricated graphene patterns: (i) micro-scale squares. (ii) micro-scale triangles. Reproduced with permission.[386] Copyright 2011, Royal Society of Chemistry. (c) SEM image of EBL patterned squares on a monolayer $MoS_2$ flake. Reproduced with permission.[388] Copyright 2020, Royal Society of Chemistry. (d) SEM image of a patterned square antidot array on a BP flake. Reproduced with permission.[389] Copyright 2017, American Chemical Society. (e) SEM images of perovskites patterns: (i) a fabricated perovskite microdisk. Reproduced with permission.[390] Copyright 2017, Wile-VCH. (ii) a perovskite 2D photonic crystals (inset shows a top view of the calculated electric field distributions). Reproduced with permission.[391] Copyright 2019, American Chemical Society. (f) Fabricated patterns on MXene-MOF films: (i) EBL patterned nanoribbons. (ii) UV-photolithography patterned concentric square ring. Reproduced with permission.[253] Copyright 2020, American Chemical Society. (g) Schematic illustration of patterning processes via lithography & lift-off. (h) Optical images of grid-patterned graphene. Reproduced with permission.[178] Copyright 2017, American Chemical Society. (i) Optical image of a square patterned GO film on a doped silica MRR. Reproduced with permission.[196] Copyright 2020, Wile-VCH. (j) Optical image of a $MoS_2$ letter pattern on $Si/Si_3N_4$ substrate. Reproduced with permission.[392] Copyright 2017, Wile-VCH. (k) Fluorescence microscopy image of perovskite interdigitated electrode patterns. Reproduced with permission.[246] Copyright 2020, American Chemical Society.

**Fig. 6(b)** shows scanning electron microscope (SEM) images of graphene patterns fabricated by lithography & etching. Both EBL and ultraviolet lithography were used to fabricate large-area graphene patterns with a feature size down to 350 nm.[386] **Fig. 6(c)** shows EBL patterned $MoS_2$ monolayers, where vapor-deposited ice was used as an EBL resist, thus allowing for contamination-free lithography.[388] **Fig. 6(d)** shows a square antidot array patterned on a BP flake with superlattice constants and radii of ~ 65 nm and ~13 nm, respectively. The antidot array of BP was demonstrated to have tunable and spatially dependent electronic properties.[389] Many perovskite devices were also fabricated, such as microdisks[390] and 2D photonic crystal cavities (**Fig. 6(e)**),[391] with feature sizes ranging from hundreds of nanometers to several microns, forming the basis of on-chip micro-lasers and solar cells. In addition, ultraviolet photolithography & plasma dry-etching has been employed to pattern MXene-MOF films (**Fig. 6(f)**), which were used to implement high performance electrical double-layer transistors.[253]

**Fig. 6(g)** shows the typical flow for lithography & lift-off processes. A resist layer is first spin-coated onto a target substrate and then patterned using EBL or photolithography. Next, a 2D film is conformally coated onto the patterned resist. Following this, the sample is immersed in a solvent that dissolves the resist and thus effectively lifts off the 2D film that is deposited on the resist, leaving only the patterned 2D film behind that is attached onto the target chip substrate. This lift-off method is widely used for fabricating metal electrodes of integrated devices, and requires both good film attachment to the resist as well as strong film adhesion to the substrate. Therefore, it is more suitable for chemically or physically bonded 2D films rather than simple physical contact. For example, self-assembled 2D films have a much better lift-off outcome than mechanically transferred 2D films.

A range of patterned 2D films using lithography & lift-off processes are shown in **Figs. 6(h) – (k)**. Grid-patterned graphene (**Fig. 6(h)**) with a width of 200 μm was fabricated via ultraviolet photolithography and lift-off,[178] showing uniform and smooth edges, with excellent electrical and optical properties. **Fig. 6(i)** shows a MRR integrated with a 50-μm-long patterned GO film,[196] where the film was deposited by solution-based self-assembly and patterned via EBL and lift-off. **Figs. 6(j)** and **(k)** show patterned $MoS_2$ and perovskite films using EBL and ultraviolet photolithography, respectively.[246, 392] The former was used for electronic devices and the latter for micro-LEDs.

***3.3.3. Nanoimprinting and pre-patterning***

Nanoimprinting (**Fig. 7(a)**) is another popular technique for integrated device fabrication in order to achieve high-resolution patterning,[393] and has recently been applied to 2D materials.[394-396] It has similarities with lithography in that resist coating, film attachment, and resist removal are all involved. The main difference is that an imprint mould is employed to pattern the resist layer instead of using EBL or photolithography, and different moulds are required to pattern different shapes. This makes it more suitable for fabricating relatively simple and repetitive patterns.

Nanoimprinting has been highly successful in fabricating graphene patterns such as nanoring arrays (**Fig. 7(b)**) and microdisks (**Fig. 7(c)**).[394, 395] By properly designing the mould and etching parameters, graphene nano-meshes with sub-10 nm ribbon widths were fabricated (**Fig. 7(d)**).[397] **Fig. 7(e)** shows nanoimprinted GO lines featuring widths of around 790 nm,[396] where a silicon mould was used, yielding low edge roughness. Recently, scalable nanoimprint patterning was used to fabricate $MoS_2$ nanoribbons, where ink precursors were used to deposit crystallized $MoS_2$ in-situ onto substrate regions defined by a PDMS stamp mould, achieving widths of 320 nm and 600 nm (**Figs. 7(f) and (g)**).[398] In addition, similar PDMS mould assisted nanoimprinting has been used to fabricate 2D perovskite patterns, including submicron square hole arrays (**Fig. 7(h)**) and micro hexagonal periodic circles (**Fig. 7(i)**).[249, 399]

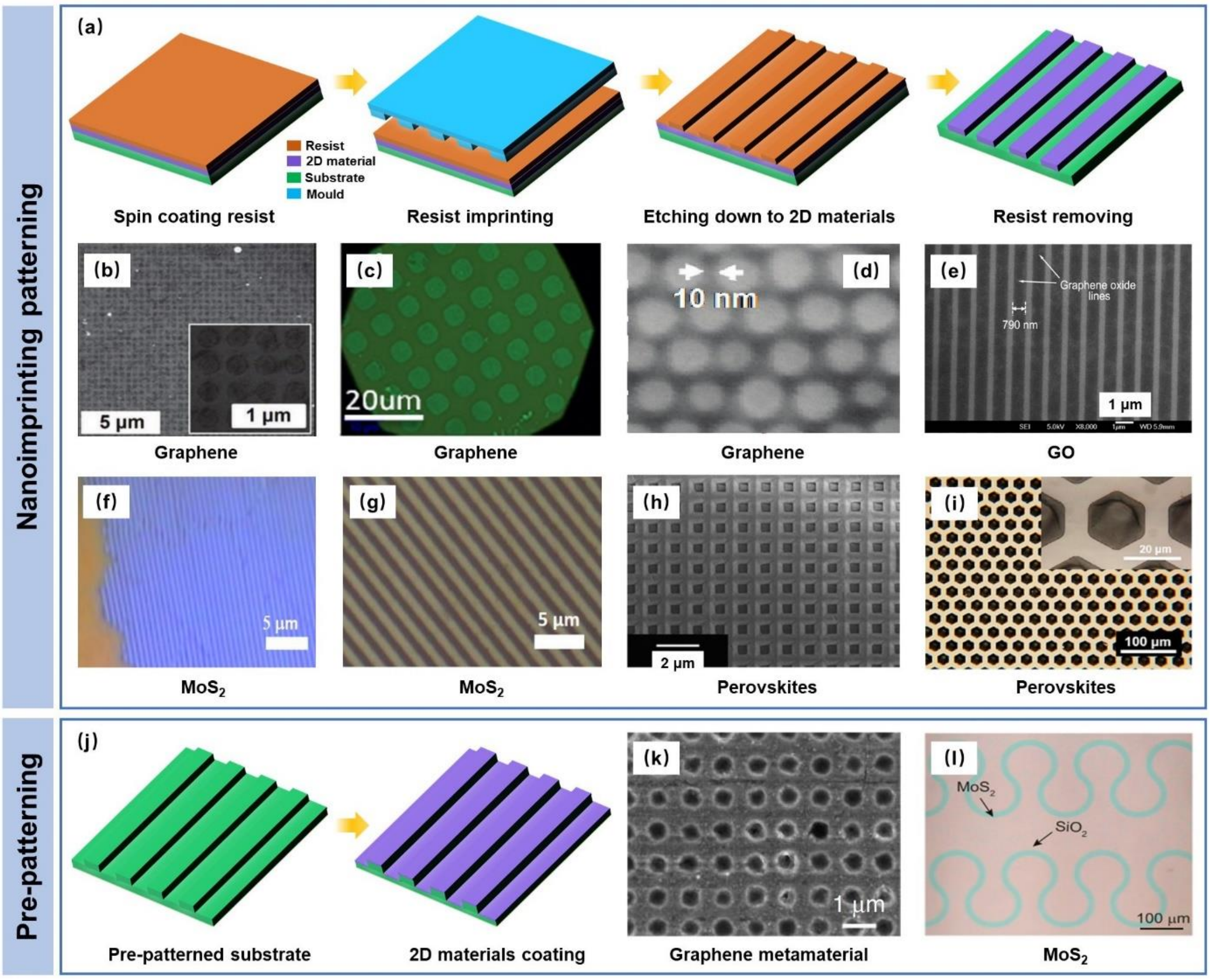

**Figure 7.** Nanoimprinting and pre-patterning techniques for 2D film patterning. (a) Schematic illustration of nanoimprinting processes. (b) SEM image of patterned graphene nanoring array. Reproduced with permission.[394] Copyright 2013, Wile-VCH. (c) Optical image of patterned graphene discs. Reproduced with permission.[395] Copyright 2013, Elsevier Ltd. (d) SEM image of graphene nanomesh. Reproduced with permission.[397] Copyright 2010, American Chemical Society. (e) SEM image of patterned GO lines. Reproduced with permission.[396] Copyright 2011, American Vacuum Society. Optical images of $MoS_2$ patterns on (f) Au thin film and (g) graphene-coated glass substrate. Reproduced with permission.[398] Copyright 2016, American Chemical Society. (h) SEM image of patterned $MAPbBr_3$ square holes. Reproduced with permission.[399] Copyright 2017, Royal Society of Chemistry. (i) Optical image of patterned hexagons periodic circles of $MAPbBr_3$. Reproduced with permission.[249] Copyright 2016, American Chemical Society. (j) Schematic illustration of patterning process with a pre-patterned substrate. (k) SEM image of a hole array of graphene metamaterial. Reproduced with permission.[400] Copyright 2020, Springer Nature. (l) Optical image of a monolayer $MoS_2$ wave array. Reproduced with permission.[401] Copyright 2019, National Academy of Sciences.

Like nanoimprinting, the direct coating of 2D materials onto pre-patterned structures (**Fig. 7(j)**) is a simple method capable of realizing large-area patterning, which also requires pre-fabrication to pattern the target substrates. Recently, this technique was used to fabricate structured solar absorbers composed of an array of holes in graphene metamaterial on a pre-patterned copper substrate (**Fig. 7(k)**).[400] The pre-patterned copper substrate was fabricated via DLW with hole diameters of ~450 nm and a period of ~600 nm. Arbitrary 2D patterns were also fabricated by directly growing $MoS_2$ via CVD on a substrate pre-patterned via photolithography followed by $O_2$ plasma treatment.[401] **Fig. 7(l)** shows the patterned monolayer $MoS_2$ wave array for FETs,[401] with a width of ~20 μm and a high film uniformity over a large area.

### *3.3.4. Other patterning techniques*

Other patterning methods include inkjet printing, focused ion beam (FIB) milling, scanning probe lithography (SPL), and self-assembled-mask lithography (SAML).

Inkjet printing is both a transfer method and an efficient patterning technique. It allows for rapid and in-situ patterning, and is particularly useful for fabricating large-area patterns with relatively low resolution (typically > 1 μm),[254, 402] where the position and pattern shape can be program controlled.[348, 402] **Fig. 8(a-ii)** shows various MXene patterns fabricated via direct printing MXene ink for micro-supercapacitors.[254] Recently, water-based 2D crystal inks were developed to fabricate large-area 2D patterns and heterostructures.[19] **Fig. 8(a-iii)** shows a large-area $MoS_2$ pattern, where different contrasts of printed regions correspond to different printing cycles. In addition, large-area arrays of photosensors and logic memory devices based on printed graphene and 2D TMDCs have also been demonstrated.[19, 402]

Focused ion beam (FIB) milling is widely used for fine-structure fabrication in the semiconductor industry, and can also be used to pattern 2D materials. Similar to DLW, FIB milling is a one-step patterning technique and free of any masks or chemical resists. Heavier ions, such as $He^+$ and $Ga^+$, are usually involved in the FIB patterning process.[403-405] To improve pattern efficiency and minimize peripheral damage, gas precursors, for example $XeF_2$, can be employed in the patterning process to effectively create a hybrid etching-assisted milling process (**Fig. 8(b-i)**). Compared to DLW and lithography, the patterning resolution of FIB is much higher. **Fig. 8(b-ii)** shows $WSe_2$ nanoribbons fabricated by $XeF_2$-assisted FIB milling.[403] The fabricated nanoribbons had an ultra-small width of 9.5 nm. By using this FIB milling method, $Mn_2O_3$ nanoribbons with a width of 10 nm (**Fig. 8(b-iii)**) were also fabricated.[406] Recently, FIB patterning was employed to tune the second-harmonic generation (SHG) in $MoS_2$ monolayers.[404] A $MoS_2$ grating structure (**Fig. 8(b-iv)**) was fabricated via focused $Ga^+$ milling, resulting in a tunable diffraction of the SHG response by adjusting the structure period.

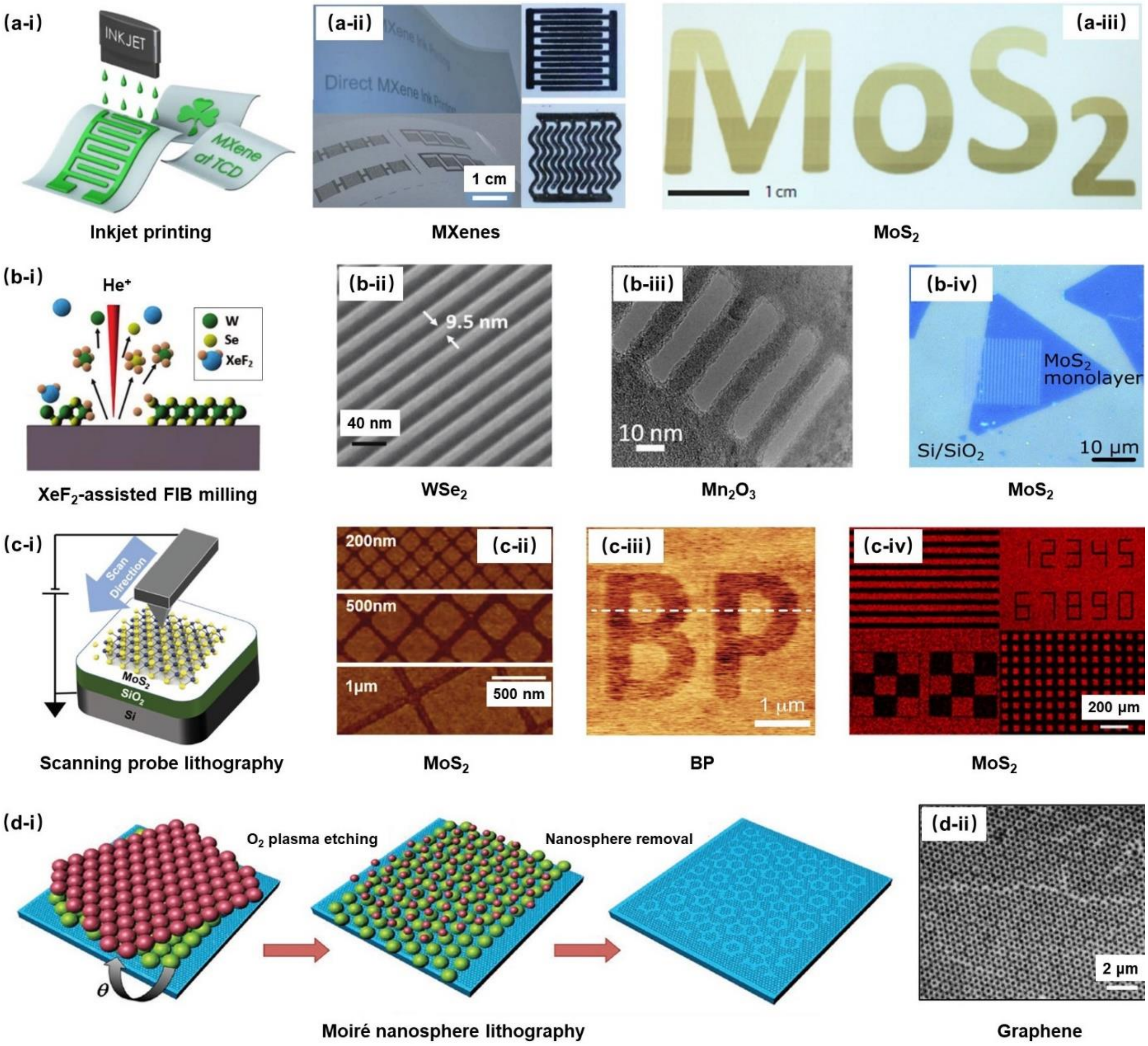

**Figure 8.** Other 2D material patterning techniques. (a) Inkjet printing: (i) schematic illustration of patterning process. (ii) images of different printed MXene patterns. Reproduced with permission.[254] Copyright 2019, Springer Nature. (iii) printed '$MoS_2$' with water-based $MoS_2$ ink. Reproduced with permission.[19] Copyright 2017, Springer Nature. (b) Focused ion beam (FIB) patterning: (i) schematic illustration of $XeF_2$-assisted FIB milling. (ii) SEM image of patterned $WSe_2$ nanoribbons. Reproduced with permission.[403] Copyright 2017, Wiley-VCH. (iii) scanning transmission electron microscope (STEM) image of patterned crystalline $Mn_2O_3$ nanoribbons. Reproduced with permission.[406] Copyright 2015, American Chemical Society. (iv) optical image of a patterned grating on monolayer $MoS_2$. Reproduced with permission.[404] Copyright 2019, OSA Publishing. (c) Scanning probe lithography (SPL): (i) schematic illustration of SPL process. (ii) atomic force microscope (AFM) images of patterned $MoS_2$ monolayer grids of different dimension. Reproduced with permission.[407] Copyright 2019, Wiley-VCH. (iii) patterned initials "BP" on a BP thin flake. Reproduced with permission.[408] Copyright 2017, Wiley-VCH. (iv) fluorescence microscopy images of $MoS_2$ films with different patterns. Reproduced with permission.[409] Copyright 2020, IOP Publishing. (d) Moiré nanosphere lithography: (i) schematic illustration of patterning process. (ii) SEM image of a fabricated graphene moiré metasurface. Reproduced with permission.[410] Copyright 2016, Wiley-VCH.

Scanning probe lithography (SPL) is another direct patterning technique that uses a scanning probe tip to fabricate micro or nanoscale patterns on 2D material films (**Fig. 8(c-i))**, where tip-induced local anodic oxidation in undesired regions of a target film and subsequent removal of the oxidation by-products are the two main steps.[407, 408] Aside from oxidation, other mechanisms can also be employed, such as thermal reduction or other chemical reactions.[411, 412] **Fig. 8(c-ii)** shows $MoS_2$ grid patterns with dimensions ranging from 200 nm to 1 μm, reflecting a high degree of dimensional flexibility. In addition, SPL patterning of BP (**Fig. 8(c-iii)**) has been achieved,[408] where transistors formed by patterned BP flakes exhibited enhanced current on–off ratios compared to pristine devices. Recently, a direct tip scratching approach[409] was utilized to achieve wafer scale patterning on graphene, $MoS_2$, and h-BN. **Fig. 8(c-iv)** shows the result in large-area $MoS_2$ films.

Another technique used to fabricate nanoscale patterns of 2D materials is self-assembled-mask lithography (SAML), and **Fig. 8(d-i)** shows an illustration of this method as applied to self-assembled moiré nanosphere lithography.[410] Here, two layers of self-assembled polystyrene nanospheres were deposited onto a graphene film to form a moiré pattern mask. An $O_2$ plasma etching was then employed to remove undesired materials and leave the pattern on the substrate. The fabricated graphene moiré metasurface (**Fig. 8(d-ii))** had a resolution of about 200 nm. Self-assembled block copolymers (BCP) lithography has also been used for patterning 2D materials. With appropriate control, BCP can self-assemble into various nanostructures, such as cylinders, spheres, vesicles, and lamellae.[413] By using BCP lithography, sub-10 nm graphene nanoribbons[414, 415] and $MoS_2$ nanorods with a width down to 4 nm have been successfully fabricated.[416]

**Table 5.** Comparison of different 2D material patterning techniques. DLW: direct Laser writing; EBL: electron beam lithography; FIB: focused ion beam; SPL: scanning probe lithography; SAML: self-assembled-mask lithography.

| Technique | Typical resolution | Speed | Prefabrication [a)] | Resist | Etching free | Industrial potential | Refs. |
|---|---|---|---|---|---|---|---|
| DLW | > 300 nm | Moderate | No | No | Yes | Moderate | [380, 417] |
| EBL | < 100 nm | Slow | No | Yes | No | Low | [389, 418] |
| Photolithography | — [b)] | Fast | Yes | Yes | No | High | [253, 386] |
| Nanoimprinting | < 10 nm | Moderate | Yes | Yes | No | High | [394, 397] |
| Pre-patterning | > 200 nm | Fast | Yes | No | Yes | High | [400, 401] |
| Inkjet printing | > 1 μm | Fast | No | No | Yes | High | [254, 402] |
| FIB | < 10 nm | Slow | No | No | Yes | Moderate | [403, 406] |
| SPL | < 100 nm | Moderate | No | No | Yes | Low | [408, 412] |
| SAML | < 10 nm | Moderate | No | No | No | Low | [414, 416] |

[a)] This includes prefabricated masks, moulds, and substrates.
[b)] Depending on the light wavelength used for lithography, the patterning resolution of photolithography ranges from several nanometres to several hundreds of nanometres.

**Table 5** compares the different 2D material patterning techniques, highlighting their unique characteristics and capabilities such as resolution and speed. Inkjet printing has the poorest patterning resolution (typically > 1 μm), followed by DLW and pre-patterning with moderate resolution on the order of hundreds of nanometers, and then EBL and SPL, and finally nanoimprint, FIB, and SAML having the highest resolution, all capable of achieving features < 100 nm in size. This high patterning resolution, however, usually comes at the expense of slow speed, which ultimately limits the area size. Techniques based on prefabricated masks, moulds, or substrates, such as photolithography, nanoimprint, and pre-patterning, typically excel at large scale, or mass production of repetitive patterns, and are thus of interest for industry applications. For DLW, FIB, and SPL, patterning and etching can be performed simultaneously, thus reducing cost and increasing efficiency. For industrial scale production, fabrication simplicity and efficiency are typically more important than resolution, while for applications requiring a high resolution, the trade-offs between resolution and efficiency need to be appropriately balanced.

### 3.4. Material property tuning / modification

Both dynamic and non-dynamic tuning or modification of the properties of 2D materials are important for enabling new functionalities and optimizing device performance. Dynamic tuning is the basis of many devices such as modulators and switches [120, 121]. Strategies for tuning or modifying the properties of 2D materials can be categorized as to whether they occur before or after on-chip transfer. Modification during the synthesis processes before on-chip transfer, such as ion intercalation doping during LPE and in-situ doping during the CVD growth, has been discussed in Section 3.1. Here, we focus on techniques implemented after transferring 2D films onto integrated chips, including gate tuning, optical tuning, strain tuning, laser modification, ion modification, molecular doping, tip modification, and other methods.

#### *3.4.1. Gate tuning*

Some 2D materials with high carrier densities and electrical conductivities are susceptible to external electrical fields, and this has underpinned important and successful approaches to electrostatic gate tuning of their electrical and optical properties, in a dynamic, reversible, and impurity-free manner.[288, 419, 420]

Ion gel or liquid gating is a popular method to tune the properties of graphene and TMDCs in integrated devices. By using an ionic liquid gate, reversible tuning of the carrier density (up to ~30%) in bilayer $NbSe_2$ was achieved.[421] A gate-tunable NLO response, including enhanced third-harmonic generation (THG) and FWM, have been demonstrated in graphene via ion gel gate tuning of its doping level and resonant conditions (**Fig. 9(a-i)**).[120] In addition, a gate-tunable metal–insulator phase transition in monolayer $MoTe_2$ was observed, resulting in significant changes in both its electrical and optical properties.[419]

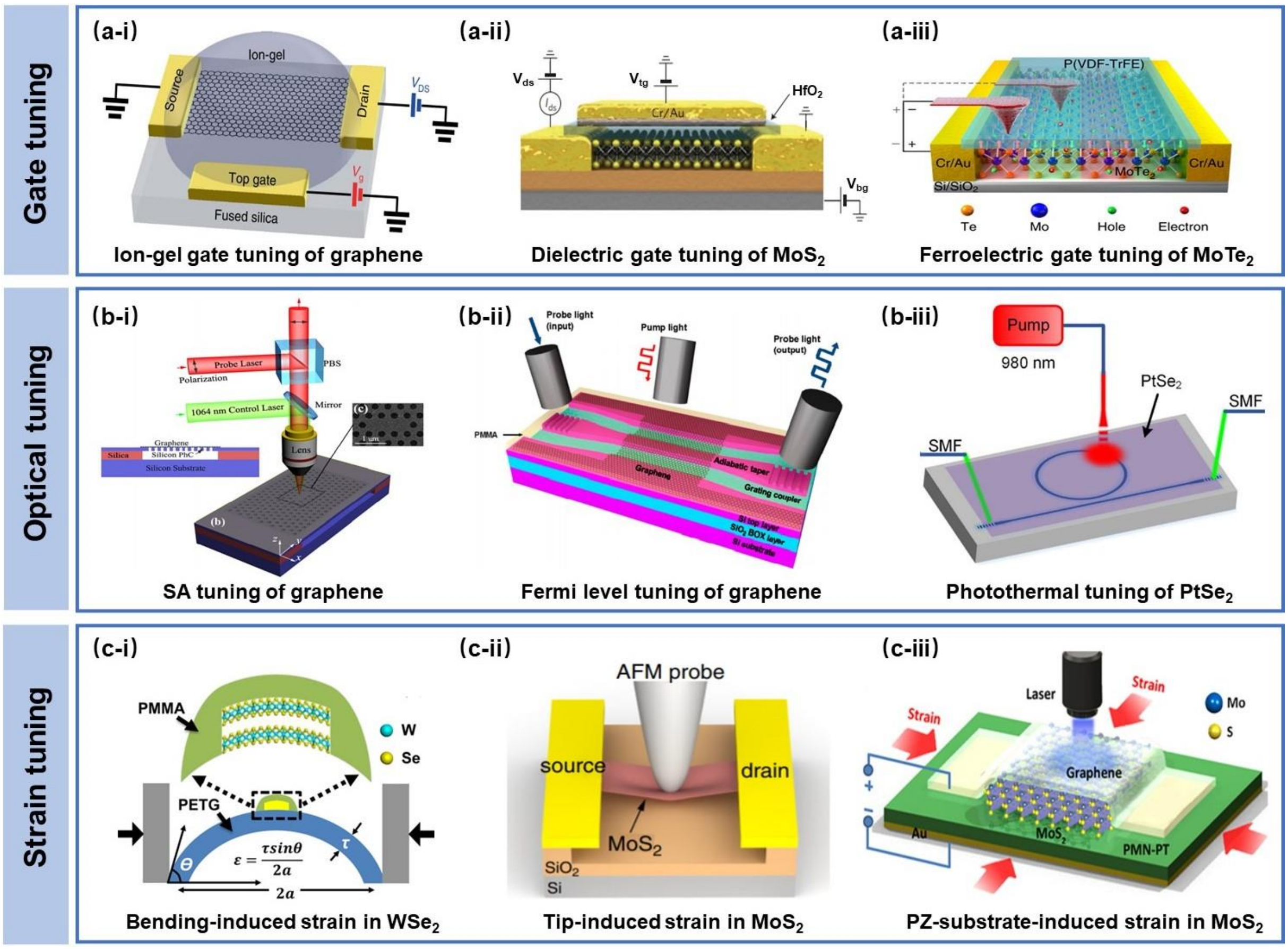

**Figure 9.** Post-treatment techniques for dynamic tuning the properties of 2D materials. (a) Gate tuning method: (i) NLO response tuning in graphene via an ion-gel gate. Reproduced with permission.[120] Copyright 2018, Springer Nature. (ii) carrier mobility engineering of $MoS_2$ with a $HfO_2$ top-gate dielectric. Reproduced with permission.[422] Copyright 2013, Springer Nature. (iii) ferroelectric gate tuning of $MoTe_2$. Reproduced with permission.[228] Copyright 2020, Springer Nature. (b) Optical tuning method: (i) saturable absorption (SA) tuning of graphene. Reproduced with permission.[274] Copyright 2015, American Chemical Society. (ii) Fermi level tuning of graphene. Reproduced with permission.[423] Copyright 2014, American Chemical Society. (iii) photothermal tuning of $PtSe_2$. SMF: single-mode fiber. Reproduced with permission.[224] Copyright 2020, OSA Publishing. (c) Strain tuning method: (i) bending-induced strain in $WSe_2$. Reproduced with permission.[424] Copyright 2014, American Chemical Society. (ii) tip-induced strain in $MoS_2$. Reproduced with permission.[425] Copyright 2015, American Chemical Society. (iii) piezoelectric (PZ)-substrate-induced strain in $MoS_2$. Reproduced with permission.[426] Copyright 2013, American Chemical Society.

In addition to ion gel gates, dielectric gates have been used to modify the carrier density and mobility in 2D materials.[422, 427, 428] **Fig. 9(a-ii)** shows a monolayer $MoS_2$ FET with a $HfO_2$ top tuning gate.[422] Compared with a single $SiO_2$ bottom-gate, the dual-gate structure enables a stronger electrostatic doping of monolayer $MoS_2$ due to the higher dielectric constant of the $HfO_2$ layer, thus yielding a large charge carrier density tuning range of over ~$3.6 \times 10^{13}$ $cm^{-2}$. By employing $HfO_2$ and h-BN-amorphous fluoropolymer covered bottom gate structures, gate tunable p-n diode and dielectric-defined lateral heterojunction were also realized in monolayer $WSe_2$ and $MoS_2$, respectively.[428, 429]

Ferroelectric gates enable property tuning in a non-volatile and switchable polarization domain.[228, 430, 431] By coupling to a pre-polarized $LiNbO_3$ substrate, a spatial carrier density modulation in graphene was demonstrated.[430] A ferroelectric polymer layer was also employed to manipulate n- and p-doping of TMDCs such as $MoS_2$[431] and $MoTe_2$.[228] **Fig. 9(a-iii)** shows a schematic of an $MoTe_2$ FET structure with a ferroelectric top layer. By using a scanning probe technique to control the polarization of ferroelectric layers, an $MoTe_2$ lateral p–n–p–n homojunction could be defined. By using a similar device structure, ferroelectric tunable PL in $WS_2$ was also demonstrated.[228]

### *3.4.2. Optical tuning*

Under laser irradiation, optical processes such as saturable absorption (SA), photon-excited carrier transport, and photothermal effects can be excited in 2D materials. These effects can be utilized for dynamic tuning of their properties such as light absorption and refractive indices.

As discussed in Section 2.1, many 2D materials exhibit strong SA,[274, 423, 432] which results in reduced light absorption with increasing power. Owing to their broadband nature and the ultrafast response of SA, broadband ultrafast optical tuning can be realized. In Ref.[274], a CW laser at 1064 nm was used to tune the optical absorption of graphene on a silicon photonic cavity at 1550 nm (**Fig. 9(b-i)**), resulting in a ~20 % variation in the quality (Q) factor. Recently, femtosecond laser pulses at 800 nm were also employed to tune the light absorption of SnSe at 633 nm,[433] achieving a modulation depth up to 44 %.

In addition to SA, photo-excited carrier transport can also be used to tune the light absorption of 2D materials. **Fig. 9(b-ii)** shows a graphene-silicon hybrid waveguide.[423] When illuminated by a 635-nm CW pump laser, the photon-excited carriers in silicon were injected into graphene through the Schottky diode junction and resulted in a carrier concentration enhancement as well as Fermi level tuning, which changed the light absorption at 1560 nm. The minimum required pump power was 2 $W/cm^2$, which was much lower than what is typically needed for optical tuning based on SA.

Optical tuning of the refractive indices of 2D materials by changing the carrier densities or employing photothermal effects has also been demonstrated.[434, 435] The changes in the refractive indices lead to phase shifts of propagating light.[434, 435] By employing a 980-nm CW light to tune the refractive indices, phase shifts up to 21 π and 6.1 π near 1550 nm were achieved in graphene and $WSe_2$ covered fibres, respectively.[435, 436] Similarly, a tuning efficiency of 0.0044 $\pi \cdot m \cdot W^{-1}$ was achieved for a $PtSe_2$/silicon hybrid MRR[224] (**Fig. 9(b-iii)**). In addition, optical tuning of THG in graphene and $MoS_2$ was also demonstrated.[437]

### *3.4.3. Strain tuning*

Strain engineering has been another effective way to tune the properties of 2D materials,[424, 438-440] although it is relatively slow compared to the gate or optical tuning. The ability to apply continuous and reversible strain to 2D layers is highly useful for dynamically tuning their properties.

By using a flexible substrate, it is possible to apply strain stress to 2D crystals via bending or stretching.[424, 441] **Fig. 9(c-i)** shows an illustration of a two-point bending apparatus for strain engineering in $WSe_2$. The substrate bending produced a uniaxial tensile strain, which induced a transition from indirect to direct bandgap together with enhanced PL in multilayer $WSe_2$.[424] By encapsulating monolayer 2D crystals in a flexible polyvinyl alcohol (PVA) substrate, efficient strain-induced bandgap modulation in $MoS_2$ and $WS_2$ has also been achieved, with a maximum modulation depth of ~ 300 meV.[441]

In contrast to uniaxial strain manipulation, tip pressing and bulging methods can produce biaxial strain in 2D films.[425, 440] By using an atomic force microscope (AFM) tip (**Fig. 9(c-ii)**), continuous and reversible strain-induced bandgap tuning in suspended $MoS_2$ was demonstrated,[425] achieving a high piezoresistive gauge factor comparable to silicon strain sensors. Recently, tunable electrical and excitonic properties via AFM tip strain-manipulation of naturally formed wrinkles in monolayer $WSe_2$ have also been reported.[439] Bulging techniques produce strain based on the deformation of 2D materials induced by the pressure difference between two sides of the material. By using this method, a strain tunable Raman response and PL were observed in graphene and $MoS_2$, respectively.[442, 443]

Piezoelectric (PZ) substrates have also been used to apply strain to graphene and TMDCs.[426, 444, 445] When applying a bias voltage perpendicular to the PZ substrate, biaxial in-plane strain can be induced in the substrate and then transferred to 2D materials. **Fig. 9(c-iii)** shows an electro-mechanical device used to apply in-plane strain to $MoS_2$,[426] generating a uniform compressive strain of up to 0.2% and resulting in tunable PL. Similarly, a strain-tunable Raman response in graphene and single photon emission in $WSe_2$ monolayers have also been reported.[444, 445] In addition to PZ techniques, microheater actuators have been employed to engineer strain in 2D materials,[446] where the mismatch of thermal expansion between the substrate and 2D films was used to create strain and thus tune the material properties.

#### *3.4.4. Laser modification*

Under high-intensity laser irradiation, the strong light-matter interaction in 2D materials can modify their properties. A laser beam focused to a small spot size on the order of microns, can drastically modify the local properties of 2D films.

Laser thinning is an efficient approach to modify the properties of 2D films via control of their thicknesses.[371, 447] For example, enhanced PL was achieved by laser thinning $MoS_2$ multilayer flakes to monolayers.[447] Laser-induced reduction of GO is widely used to tune its electrical and optical properties (**Fig. 10(a-i)**). Bandgap tuning of GO from 2.1 eV to 0.1 eV was achieved by using femtosecond laser pulses.[44] Tunable optical absorption and PL were also demonstrated.[44, 123, 417] The NLO response of GO can also be modified – both the transition of the NLO absorption response from SA to reverse SA and Kerr nonlinearity from positive to negative have been achieved during laser reduction.[124]

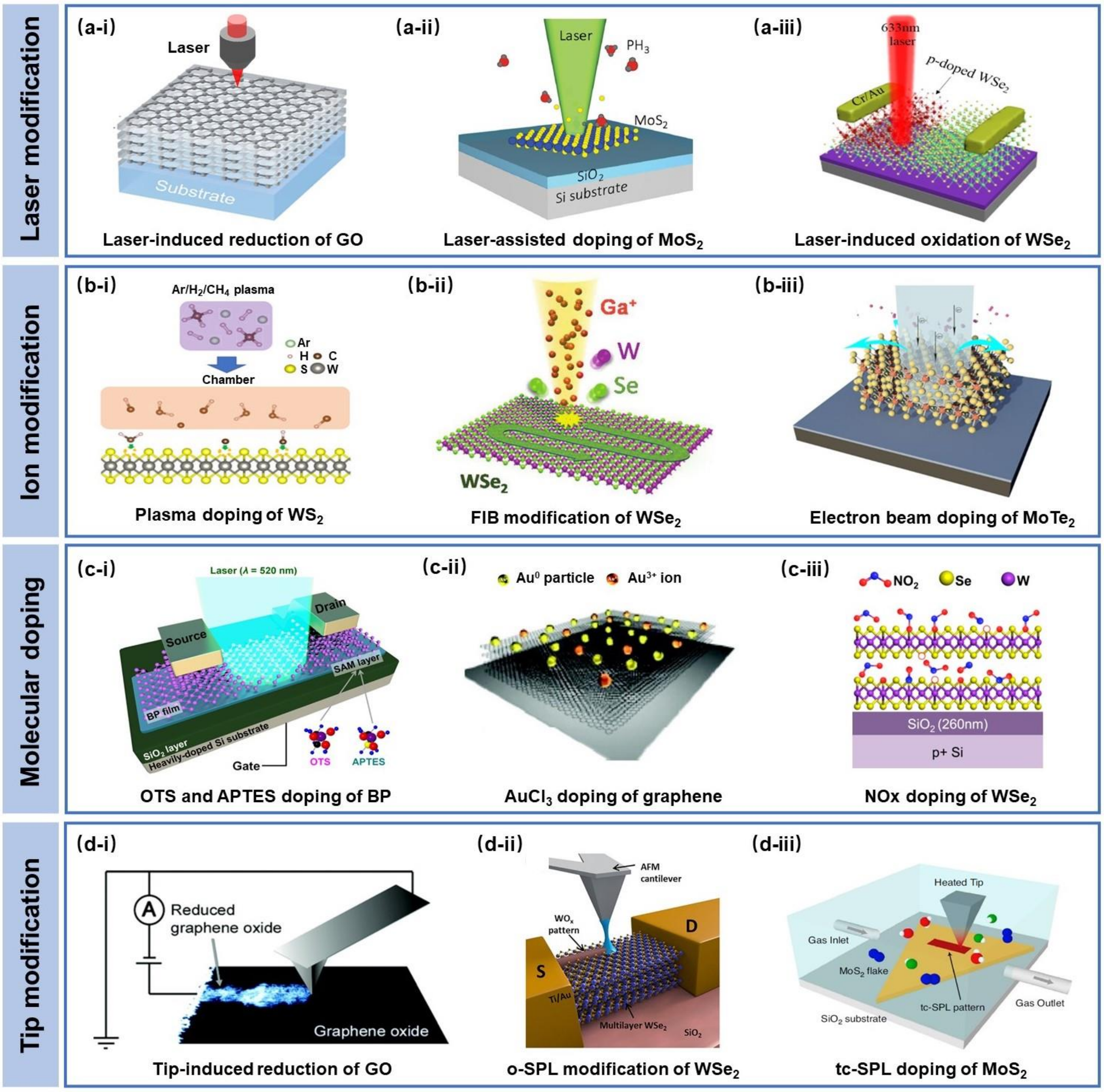

**Figure 10.** Post-treatment techniques for modifying the properties of 2D materials. (a) Laser modification method: (i) laser-induced reduction of GO. Reproduced with permission.[44] Copyright 2019, American Chemical Society. (ii) laser-assisted doping of $MoS_2$. Reproduced with permission.[211] Copyright 2015, Wiley-VCH. (iii) laser-induced oxidation of $WSe_2$. Reproduced with permission.[213] Copyright 2019, American Chemical Society. (b) Ion modification method: (i) plasma-assisted doping of $WS_2$. Reproduced with permission.[448] Copyright 2019, AAAS. (ii) defect engineering in monolayer $WSe_2$ via focused $Ga^+$ ion beam irradiation. Reproduced with permission.[449] Copyright 2020, Royal Society of Chemistry. (iii) electron beam induced doping in multi-layer $MoTe_2$. Reproduced with permission.[229] Copyright 2020, Elsevier B.V. (c) Molecular doping method: (i) octadecyltrichlorosilane (OTS) and 3-amino-propyltriethoxysilane (APTES) assisted doping of BP. Reproduced with permission.[450] Copyright 2017, American Chemical Society. (ii) $AuCl_3$ doping of graphene. Reproduced with permission.[451] Copyright 2010, American Chemical Society. (iii) NOx doping of $WSe_2$. Reproduced with permission.[452] Copyright 2014, American Chemical Society. (d) Tip modification method: (i) tip-induced reduction of GO. Reproduced with permission.[453] Copyright 2015, Elsevier Ltd. (ii) o-SPL treatment of $WSe_2$. Reproduced with permission.[454] Copyright 2016, AIP Publishing. (iii) tc-SPL doping of $MoS_2$. Reproduced with permission.[411] Copyright 2020, Springer Nature.

Laser-assisted doping has achieved site-specific selectivity and hence property tuning of 2D materials.[211, 455] **Fig. 10(a-ii)** shows an illustration of laser-assisted doping in $MoS_2$ with a $PH_3$ dopant precursor. The focused laser beam simultaneously induces sulphur vacancies in $MoS_2$ and the dissociation of dopant molecules incorporated into the vacancy sites. Based on this, phosphorus doping in $MoS_2$ and $WSe_2$ has been achieved, which induced a suppression of n-type characteristics in $MoS_2$ as well as an enhancement of p-type characteristics in $WSe_2$.[211] By using an $H_2S$ precursor, sulphur doping in monolayer $MoSe_2$ and $WSe_2$ has also been achieved, together with the modification of their PL and electrical properties.[455]

Laser-induced oxidation and phase transitions can also modify the properties of 2D crystals. Two-photon oxidation was used to modify the electrical and optical properties of graphene,[456] while a tunable bandgap in BP flakes was also demonstrated via laser-induced oxidation.[135] Recently, a laser scanning technique was used to fabricate a lateral $WSe_2$ p–n junction (**Fig. 10(a-iii)**). It was found that the laser-oxidized product $WO_x$ was responsible for the p-doping in $WSe_2$.[213] Laser-induced phase transitions have also been demonstrated in 2D $MoTe_2$,[457] $MoS_2$,[458] and $PdSe_2$.[340] Based on the 2H to 1T phase transition, enhanced carrier mobilities and modified PL properties were achieved in $MoTe_2$ and $MoS_2$, respectively.[457, 458] A semiconductor to semi-metal transformation was also observed in $PdSe_2$ films under different laser intensities.[340]

### *3.4.5. Ion modification*

Similar to laser modification, energetic ion irradiation can also be used to tailor the properties of 2D materials. It mainly includes plasma-assisted doping, FIB modification, and electron beam modification.

In plasma-assisted doping, energetic ions generated from different gas molecules are employed to either create site vacancies that facilitate dopant incorporation, or directly implant them into the 2D crystal lattice. Ammonia plasma treatment was used to fabricate large-area nitrogen-doped graphene films.[459] A maximum doping level of 3 % nitrogen was achieved, resulting in a shift of 0.2 eV for the valence band maximum. Plasma surface treatment with $N_2$ was also used to achieve p-doping of $MoS_2$ via a nitrogen substitution of the chalcogen sulphur.[460] **Fig. 10(b-i)** shows a schematic illustration of a plasma-assisted carbon doping process in monolayer $WSe_2$.[448] A $CH_4$ dopant precursor mixed with $Ar/H_2$ carrier gas was employed in the doping process. The carbon doped $WSe_2$ monolayer showed a decrease in optical bandgap from 1.98 eV to 1.83 eV, along with modified electronic transport characteristics.

FIB can be used not only for patterning 2D films, but also for controllably altering the material properties.[449, 461] **Fig. 10(b-ii)** shows a schematic of $WSe_2$ modification via FIB irradiation. Focused $Ga^+$ beam scanning was used to introduce atomic defects into a $WSe_2$ monolayer. Trapped electrons/holes or defect-bound excitons were generated by the FIB irradiation, which exhibited a recombination lifetime as long as 4 μs.[449] A focused helium ion beam was also employed to tailor the optical and electrical properties of $MoSe_2$.[462] With a local defect tuning in $MoSe_2$, an enhanced PL intensity of B-exciton was achieved.

Electron beam irradiation has also been utilized for 2D material modification.[229, 463, 464] Using this method, air-stable n-doping in $MoTe_2$ was achieved (**Fig. 10(b-iii)**).[229] The doping level was controlled by changing the electron beam energy.[229] Similarly, selective defect modification in $WSe_2$ was achieved,[463] where the generated defect-bound excitons in $WSe_2$ exhibited extraordinary optical properties including a recombination lifetime approaching 200 ns and a valley lifetime longer than 1 μs. Based on electron beam doping, the precise modification of the carrier characteristics such as subthreshold swing and carrier mobility in $MoS_2$ and graphene has also been reported.[464]

### *3.4.6. Molecular doping*

The molecular doping that we discuss here refers to introducing dopants onto 2D films after their on-chip transfer. Due to the difference in Fermi level, charge transfer between the host material and adjacent dopant results in either depletion or accumulation of carrier majorities in 2D layers, thus allowing for effective doping and property modification.

Organic molecules with versatile functional groups have been widely used for 2D crystal doping via charge transfer processes. For example, triphenylphosphine ($PPh_3$) molecules were successfully used to achieve n-doping in $WSe_2$, where the phosphorus atoms in $PPh_3$ generated lone pairs of electrons that enabled the electron donation to $WSe_2$ layer at a level of $10^{11}$ $cm^{-2}$.[212] Self-assembled layers of 3-amino-propyltriethoxysilane (APTES) and octadecyltrichlorosilane (OTS) have also been employed to induce n-doping and p-doping in BP (**Fig. 10(c-i)**).[450] Further, DNA molecules and metal-modified DNA were used for n-doping and p-doping in $MoS_2$ and $WSe_2$, respectively.[465]

By decorating metals or metal compounds, 2D materials can also be doped. A range of doping effects using different metal nanoparticles have been demonstrated,[466] where metals such as Au, Ag, Pd, Pt, and Sc, led to a p-doping of $MoS_2$, whereas Y enabled a n-doping. $AuCl_3$ doping was used to tune the electrical properties of graphene (**Fig. 10(c-ii)**), yielding a decreased sheet resistance and improved environmental stability. By coating a $AuCl_3$ layer on top, a p-doing in $MoS_2$ was achieved.[467] Metal oxidation can also be used for 2D material doping. Recently, a rapid flame synthesis method was developed to deposit $MoO_3$ nanosheets onto $WSe_2$, which served as a p-doping layer to modify the sheet conduction of $WSe_2$.[468]

Gas molecules can also be used as active doping media to modify 2D materials. **Fig. 10(c-iii)** shows a schematic of a doping process in $WSe_2$ via covalent functionalization by $NO_x$.[452] An air-stable p-doping in $WSe_2$ was achieved, resulting in a contact resistance reduction between $WSe_2$ and Pd electrodes as well as a degenerated doping concentration of $1.6 \times 10^{19}$ $cm^{-3}$. By physically absorbing $O_2$ and $H_2O$ molecules, up to 100 times enhancement in the PL efficiency of n-doping $MoS_2$ was demonstrated.[469] The charge transfer between the adsorbed gas molecules and the 2D semiconductors significantly reduced the electron density and hence improved the light emission.

#### *3.4.7. Tip scanning modification*

Similar to SPL patterning, tip-induced reduction, oxidation, or other reactions can be used for localized property modification of 2D materials. Tip-induced reduction has been widely used for modifying GO.[453, 470] **Fig. 10(d-i)** shows a schematic of GO reduction in ambient atmosphere via a conductive AFM tip scanning. By applying a negative voltage, a local reduction of the GO film at the probe location was achieved.[453] Thermal probe tips were also used to reduce GO films and tune its electrical and optical properties.[471] High temperature heating by a tip, of around 1000 °C was required to realize complete GO reduction. In addition, highly efficient catalysed reduction of GO in a hydrogen atmosphere has been demonstrated by using a platinum-coated AFM tip.[472]

Anodic oxidation of 2D films induced by a scanning tip is another effective method to modify their properties. By using oxidation SPL (o-SPL), partial oxidation was employed in the gap opening and property modification of single- and bilayer graphene.[473, 474] Tip-induced $WO_x$ (**Fig. 10(d-ii)**) was used to modify $WSe_2$,[454] yielding FETs with a better subthreshold swing and a higher ON/OFF ratio. Localized anodic oxidation was also employed in modifying $MoS_2$,[475] where enhanced PL was observed in monolayer $MoS_2$ covered with an oxidation layer. In addition to graphene and TMDCs, tip-induced anodic oxidation was also employed in modification and patterning of BP.[408]

Thermochemical SPL (tc-SPL) can be used to control defects and doping in 2D materials.[411, 476] As discussed above, the thermal tip technique can be used to reduce GO in localized positions. In addition, tc-SPL can be used to realize controllable doping in 2D crystals when assisted by different active gases. **Fig. 10(d-iii)** shows a schematic of tc-SPL doping of monolayer $MoS_2$.[411] p-doping in $MoS_2$ was achieved in an HCl/$H_2O$ gas atmosphere, whereas n-doping was observed in an inert $N_2$ environment. Compared to plasma and molecular doping, tip-induced doping provides a localized material modification in a nanoscale resolution. [411]

### *3.4.8. Other tuning / modification techniques*

The intercalation of foreign ions and atoms into host materials has been used to modify 2D crystals. Recently, a solvent-based technique was developed to intercalate Cu and Co atoms in 2D $SnS_2$.[477] Combining the intercalation technique with lithography, a spatially controlled process was used to seamlessly integrate n-type and p-type semiconductors and metal in 2D materials.[477] Electrochemical techniques have also been used to realize $Li^+$ and $Na^+$ intercalations in $MoS_2$ in order to modify its electrical and optical properties.[478] By using a $LiClO_4$-polyethylene oxide solid electrolyte, gate-tunable $Li^+$ intercalation in 1T-$TaS_2$ was also achieved,[479] where a phase transition from a Mott-insulator to a metal was observed.

Thermal annealing has been used to create defects and induce doping in 2D materials. The anion vacancies generated by thermal annealing yielded an enhanced PL intensity in monolayer TMDCs.[480] By using simple thermal annealing of a GO/$B_2O_3$ mix, boron-doped graphene with a high specific capacitance was achieved.[481] Other techniques, such as flash light heating[482] and microwave treatment,[483] have been used to reduce GO films and modify their properties.

**Table 6.** Comparison of different on-chip property tuning / modification techniques for 2D materials.

| Technique | Reversibility | Response time | Additional fabrication [a)] | Large-area capability | Patterning capability | Industrial potential | Refs. |
|---|---|---|---|---|---|---|---|
| Gate tuning | Yes | Fast | Yes | Moderate | No | High | [120, 228] |
| Optical tuning | Yes | Fast [b)] | No | Low | No | Low | [224, 274] |
| Strain tuning | Yes | Moderate | Yes | Moderate | No | Moderate | [424, 426] |
| Laser modification | No | — | No | Moderate | Yes | Moderate | [44, 213] |
| Plasma-assisted doping | No | — | No | High | No | High | [448, 460] |
| FIB modification | No | — | No | Low | Yes | Moderate | [449, 462] |
| Electron beam modification | No | — | No | Low | Yes | Low | [229, 463] |
| Molecular doping | No | — | Yes | High | No | High | [450, 467] |
| Tip modification | No | — | No | Low | Yes | Low | [411, 453] |
| Ion intercalation | No | — | Yes | High | No | High | [477, 478] |
| Thermal treatment | No | — | No | High | No | Moderate | [480, 481] |

[a)] This includes fabrications of additional metal gates, flexible or PZ substrates, dopant layers, and so on.
[b)] Here indicates the response time of all-optical tuning technique. For photothermal tuning method, the response time is relatively slower.

**Table 6** compares the different tuning / modification techniques for 2D materials. Amongst them, gate tuning, optical tuning, and strain tuning allow reversible and dynamic tuning, which is attractive for devices such as modulators and switches. Other techniques, such as laser modification, FIB, and tip scanning modification, enable the localized optimization of material properties, as well as for patterning. Simultaneously combine patterning with property modification is a powerful capability that can simplify fabrication, making these techniques promising for industrial applications. For plasma-assisted doping and molecular doping, low-cost and large-area modification capability are their advantages, which are also needed in mass production.

Current property tuning or modification strategies still face challenges, with many underlying mechanisms of the processes still not being completely understood, making highly accurate material tuning or modification challenging. In addition, although some techniques have advantages for industrial applications, such as patterning via laser modification or tip scanning, as well as large-area processing via plasma-assisted and molecular doping, the uniformity and efficiency of these approaches still need to be further improved.

## 4. 2D heterostructures

Beyond employing only one material, assembling different 2D crystals into van der Waals (vdW) heterostructures offers many new features and possibilities, and there has been an enormous surge in activities in this field.[15, 138] In this section, we review and discuss the state-of-art fabrication techniques for on-chip integration of 2D heterostructures.

Generally, 2D crystals have unique characteristics that can vary even for the same material. For example, many properties are highly dependent on film thickness.[128, 132] By stacking different 2D layers together, charge redistribution, interface strain, and structural modifications in each material can be engineered, thus leading to many extraordinary phenomena.[138, 484] Excellent electrical and optical properties, such as enhanced carrier mobility,[485, 486] photovoltaic performance,[487, 488] broadband photon absorption,[489] and tunable nonlinear optical properties,[490, 491] have been demonstrated in vertical or lateral 2D heterostructures. In addition to orderly stacking manipulation, twisting adjacent 2D crystals with a small angle can give rise to unprecedented characteristics. For example, unconventional superconductivity[492], as well as a strong MIR photoresponse[493], were observed in twisted bilayers or so-called magic-angle graphene systems. Moiré-trapped valley excitons and Wigner crystal states in TMDC bilayer twisted structures have also been reported.[494, 495]

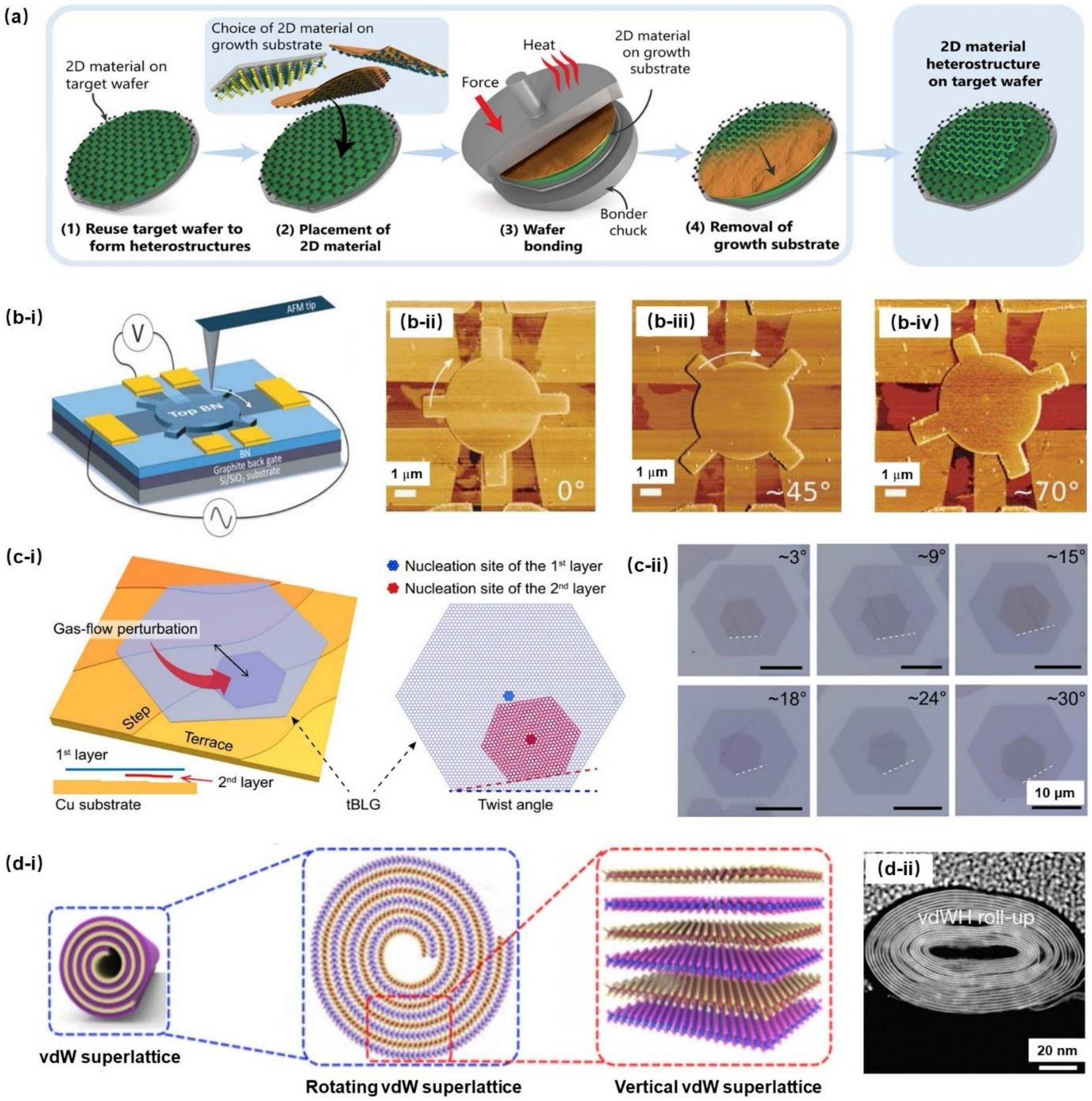

**Figure 11.** (a) Schematic illustration of wafer scale 2D heterostructure formation. Reproduced with permission.[496] Copyright 2021, Springer Nature. (b) Rotatable heterostructures: (i) schematic of the designed device. (ii) – (iv) AFM images of a fabricated device showing three different orientations of the top BN. Reproduced with permission.[497] Copyright 2018, AAAS. (c) Twisted bilayer graphene (tBLG): (i) schematic showing the hetero-site nucleation strategy for growing tBLG on a Cu substrate. (ii) optical images of tBLGs with twist angles of ~3°, ~9°, ~15°, ~18°, ~24°, and ~30°. Reproduced with permission.[498] Copyright 2021, Springer Nature. (d) High-order vdW superlattice: (i) schematic of the high-order superlattice in the vdW heterostructure roll-up. (ii) cross-sectional STEM image of a representative $SnS_2/WSe_2$ roll-up. Reproduced with permission.[499] Copyright 2021, Springer Nature.

The fabrication of 2D heterostructures, which is usually more complex than single homogeneous 2D materials, is critical for fully exploiting their properties, and artificial layer-by-layer stacking is a key approach. Early work mainly focused on conventional wet or dry transfer methods,[485-487] but it was found that the polymer residuals on top of the 2D crystals could deteriorate the device performance and hinder further stacking for the multilayers. To resolve this issue, a pick-up strategy was developed to construct 2D heterojunctions with multilayer structures and special alignments.[334, 336, 495] To meet the requirements of large-area production, a universal strategy was developed to fabricate two-layer heterostructures with a lateral size up to 100 mm.[496] **Fig. 11(a)** shows the corresponding fabrication steps, where a bottom 2D film was first prepared on a commercial integrated wafer with a bisbenzocyclobutene adhesive layer and another 2D film was then stacked onto the first one after a wafer bonding process. By using the pick-up method, rotatable BN-graphene-BN heterostructures (**Fig. 11(b-i)**) were fabricated, which allowed dynamical in-situ twisting via an AFM tip.[497] **Figs. 11(b-ii)** to **(b-iv)** show the fabricated devices with different BN orientations.

In addition to artificial stacking, direct growth methods, such as epitaxy and CVD, have also been used to fabricate 2D vdW heterostructures.[138, 488, 500] These approaches could eliminate the chemical or polymer contaminations during the stacking process and thereby enable the large-area fabrication of 2D heterojunctions. Epitaxy methods fabricate heterostructures by growing other materials on a prepared 2D crystal substrate, with graphene, h-BN, and some TMDCs being the most widely used substrates.[138] By using this approach, graphene-BN,[501] $WSe_2$-graphene,[502] and $MoSe_2$-$Bi_2Se_3$[503] stacks have been successfully fabricated. The direct growth of 2D heterostructures based on CVD methods has also been achieved.[488, 498-500] By using one- or two-step CVD approaches, vertical and lateral heterostructures, such as $WSe_2$-$MoS_2$[488] and $WS_2$-$MoS_2$,[500] have been fabricated. Further, by using hetero-site nucleation strategy, high-quality twisted bilayer graphene (tBLG) has also been fabricated (**Fig. 11(c-ii)**),[498] showing a large range of twist angles ranging from ~ 3° to 30° for engineering their properties. In addition to the flat heterostructures, high-order vdW superlattices have also been realized by rolling up synthesized heterostructures.[499] **Fig. 11(d)** shows a schematic of the high-order superlattices as well as STEM image of as-fabricated $SnS_2/WSe_2$ roll-up.

Inkjet printing can also be used to fabricate large-area heterostructures. Recent advancements of universal inks for a wide range of 2D crystals[349] as well as programmed jetting manipulation[314, 339] make it possible to print multilayer heterojunctions for on-chip applications. By directly printing BP onto graphene/Si devices, a BP-graphene-Si Schottky junction photodetector was fabricated.[314] Recently, all-printed large-area graphene-BN-$WSe_2$ heterojunctions were also fabricated by sequence printing different 2D crystal inks.[402]

Despite this significant progress, there are still challenges for on-chip integration of 2D heterostructures. First, the limitations of current fabrication techniques for on-chip integration of single 2D materials still exist, which is even more challenging when dealing with the compatibility issues of the different fabrication techniques for different 2D materials. This limits the scale and cost-effectiveness of current integrated devices incorporating 2D heterostructures. Secondly, although it has been demonstrated that extraordinary properties can be obtained by properly designing the material stacking sequence, interlayer spacing or twist angles, the accurate and stable control of these parameters with current manipulation strategies is still difficult, thus limiting the full exploitation of the advantages of 2D heterostructures. Lastly, most integrated devices incorporating 2D heterostructures either have no material tuning modules or at most have only very simple ones. The feasibility of combining a variety of tuning techniques, such as those mentioned in Section 3.4, remains an open challenge.

However, even though challenges remain, there are no fundamental roadblocks to the industrial implementation of 2D heterostructures for chip-scale integrated devices. With the family of 2D materials continuously growing, many new materials with distinctive properties are represented by 2D heterostructures. This will enrich their on-chip applications and in the meantime create new requirements for device fabrication.

## 5. Challenges and perspectives

The past decade has witnessed tremendous progress in fabrication techniques for chip-scale devices incorporating 2D materials, that have significantly improved the capability of conventional integrated chips. Despite this success, there are still challenges and new demands for future development. In this section, we discuss the current challenges and future perspectives.

In **Table 7**, we summarize the fabrication techniques for on-chip integration of 2D materials. For more well established 2D materials such as graphene and TMDCs, all of the fabrication techniques have been demonstrated, whereas for newer materials such as MXenes, MOFs, and graphdiyne, some techniques remain to be explored. In addition, the particular form of each material property affects which techniques can be applied. For example, synthesised GO and graphdiyne are usually either nanosheet powder or solution, and so are unsuitable for ME and dry transfer. For each technique in **Table 7**, the advantages, challenges, and potential for large-scale production have been discussed in Section 3. Generally, each technique has advantages for specific applications depending on the material and device structure, and so no single technique is universally better.

**Table 7.** Summary of on-chip fabrication techniques for 2D materials. ME: mechanical exfoliation. LPE: liquid phase exfoliation. CVD: chemical vapor deposition.

| Technique | Graphene | GO | TMDCs | BP | h-BN | Perovskites | MXenes | MOFs | Graphdiyne |
|---|---|---|---|---|---|---|---|---|---|
| ME | [37] | — | [47] | [158] | [137] | [324] | — | [327] | — |
| LPE | [282] | [309] | [284] | [313] | [317] | [325] | [290] | [328] | — |
| CVD | [293] | [310] | [291] | [504] | [318] | [148] | [505] | [331] | [506] |
| Dry transfer | [6] | — | [214] | [337] | [333] | [324] | — | [327] | — |
| Wet transfer | [16] | — | [204] | — | [507] | [508] | — | — | — |
| Solution dropping | [339] | [346] | [349] | [314] | [19] | [145] | [254] | [253] | [268] |
| Self-assembly | [352] | [27] | [353] | — | — | — | [351] | [509] | — |
| Laser patterning | [366] | [380] | [379] | [135] | [510] | [382] | [381] | [511] | — |
| Lithography | [386] | [196] | [388] | [389] | [493] | [246] | [253] | [253] | — |
| Nanoimprinting | [394] | [396] | [398] | — | — | [249] | — | [512] | — |
| Pre-patterning | [513] | [400] | [401] | — | — | [514] | [515] | [516] | [517] |
| Gate tuning | [120] | — | [422] | [518] | [242] | [519] | [520] | — | [268] |
| Optical tuning | [274] | [521] | [224] | [522] | — | [523] | [258] | [524] | [525] |
| Strain tuning | [443] | [526] | [424] | [527] | [528] | [529] | [530] | — | — |
| Laser modification | [456] | [44] | [211] | [135] | [510] | [531] | [532] | [533] | — |
| Ion modification | [459] | [534] | [460] | [535] | [536] | [537] | [538] | [539] | [540] |
| Molecular doping | [541] | [542] | [212] | [450] | — | [543] | [544] | — | [545] |
| Tip modification | [546] | [453] | [454] | [408] | — | — | — | [547] | — |

For large-scale production, techniques based on CVD, inkjet printing, and lithography have shown strong potential for industrial scale manufacturing, but do have limitations in terms of low film quality or production efficiency. Another limiting factor is the lack of unified synthesis protocols or fabrication standards, particularly since 2D materials are highly affected by their fabrication process. The more established materials such as graphene, GO, and TMDCs, are ready to move forward to this stage, while the newer materials still need more research.

Material stability is important for practical applications, and for many hybrid integrated devices, since the 2D films are implemented as the top functional layers, it makes the device vulnerable to environmental conditions. To avoid material deterioration and stabilize the device performance, proper packaging is needed. This is particularly true for those with poor air stability, such as BP, silicene, and monolayer TMDCs. To address this, dielectrics, such as $Al_2O_3$,[548] specific polymers, and h-BN,[549] have been employed as the packaging layers. Nevertheless, the poor interfacial quality between 2D films and $Al_2O_3$ or polymer materials,[550, 551] as well as the limited covering area of h-BN make this approach challenging for efficient wafer-scale packaging. New encapsulation materials and optimized deposition methods still need to be developed.

Temperature is another issue that influences the properties and stability of 2D materials – high temperature can significantly degrade performance or even create permanent damage. To better control the temperature, thermal dissipation structures[552] or temperature controllers[553] can be implemented. Moreover, stable and compact adhesion between 2D layers and chip substrates or electrodes is crucial for stabilizing the device performance, and this can be achieved by substrate surface modification such as rapid thermal annealing,[554] plasma treatments,[555] or by introducing assisted interfacial layers.[556] There are also other factors that can affect material stability, such as contamination or bubbles at the interfaces or internal residual stress induced by the synthesis or transfer processes. Therefore, interface cleaning via electrical current annealing,[557] plasma treatment,[558] and mechanical cleaning,[559] or high-efficient strain releasing techniques, such as post-annealing[560] and ultrasonic treatments,[561] can all be useful.

Material characterization techniques are crucial to develop fabrication processes. Mainstream techniques, such as transmission electron microscope (TEM), AFM, Raman spectroscopy, and scanning tunnelling microscope (STM), have been widely employed to identify the fundamental properties of 2D crystals, such as layer number, material defects, crystal orientation, and interfacial morphologies.[37, 131, 342, 343] By introducing external control of temperature, electrical field, or a combination of them with specific film synthesis methods, in-situ characterization techniques can be realized,[562, 563] which enables the investigation of the dynamic behaviour of film growth. These methods tend to involve high-end equipment and are suited to small samples, whereas for industry scale applications, more research is needed on fast and in-line monitoring methods.

The accurate measurement of material properties is crucial for the design and fabrication of integrated devices incorporating 2D materials, and the reverse is true – device performance can act as useful feedback for the synthesis and fabrication processes. For example, the carrier mobilities of graphene and TMDCs can be characterized by integrating them onto chips to form FETs.[6, 17] By integrating 2D layered GO films with optical waveguides, layer dependent nonlinear optical properties of the GO films can also be characterized,[28, 29] which is challenging for conventional Z-scan methods due to the weak response of ultrathin 2D films.

For integrated devices, the ability to recycle chips at the end of their life is an important consideration not just for the environment but to reduce costs. The removal and re-transfer of 2D films are critical for chip reuse, particularly given that the fabrication of integrated devices is usually more complex and costly than that of 2D films. To date, different techniques have been proposed to remove 2D materials from integrated devices. For example, by using flat polymers as a mechanical peeling medium, 2D flakes can be fully disassembled from the bottom substrates.[564] In addition, wet-chemical etching, oxygen plasma exposure, and laser ablation also allow the removal of undesired 2D films.[374] However, with respect to the removal efficiency and contamination control, current techniques still need to be optimized to satisfy the requirements of mass production, where modular designs could be instrumental. Methods such as developing independent 2D material units that can be readily integrated onto target chips, will minimize the impact of unqualified material synthesis and transfer, thus improving the production efficiency.

The integration fabrication techniques for other classes of materials can also be utilized. For example, a two-beam optical lithography technique that utilizes both focused writing and inhibition beams was developed for high-resolution patterning in a two-photon absorption (TPA) photoresin, achieving a feature size down to 9 nm.[565] Due to the strong TPA in many 2D materials, this technique could also be used for patterning 2D films, potentially increasing the resolution to below 10 nm by properly selecting the laser wavelength and beam power. Recently, a digital mask based on a digital-micromirror-device was developed for femtosecond projection via two-photon lithography for high-throughput fabrication of arbitrary structures in photopolymers, achieving sub-micrometer resolution.[566] The high-shape-flexibility and controllability of the digital mask are attractive for 2D material patterning, especially for the scalable fabrication of complex functional nanostructures. In addition, the efficient and controllable fabrication of porous nanostructures in 2D crystals is useful for energy devices and sensors. The well-developed fabrication techniques for porous silicon and metals, such as metal-assisted catalytic etching[567] and 3D addictive printing,[568] may provide new strategies for the fabrication of porous 2D materials.

There are new emerging techniques continually being introduced for the on-chip integration of 2D materials. Recently, machine learning has proven to be a powerful technique for material prediction, property analysis, and device optimization.[569] This technology can not only offer property-oriented 2D material design for a given application, but also be used for multi-dimensional analysis and fast fabrication evaluation. DNA origami is a technique that supports the fabrication of complicated 2D and 3D nanostructures ranging from several nanometres to sub-micrometres, which is used for biosensors, catalysis, and nanofabrication.[570] Using DNA origami structures as nano-templates or masks, combined with in-situ material synthesis or film transfer, can enable the fabrication of functional 2D film patterns potentially down to the sub-5 nm level.

The on-chip integration of 2D materials represents an exciting and still evolving field, residing at the intersection between integrated devices and 2D material science. Integrated device platforms with well-developed fabrication techniques can provide strong support for 2D materials to unlock their extraordinary properties. 2D materials, in turn, offer significant opportunity to improve conventional integrated devices. Along with the expansion of the 2D material family itself, the advances in fabrication techniques to incorporate 2D materials onto chips offers significant promise to realize unprecedented capabilities of future devices. This is particularly the case when these techniques are combined with CMOS fabrication techniques for nonlinear photonic chips including microcombs [571-620].

## 6. Conclusion

Born from the marriage of integrated devices and 2D materials, the on-chip integration of 2D materials represents a new frontier for implementing functional devices for a great many applications. In the past decade, significant advances in this field have been made by virtue of the development of device fabrication techniques. In this review, we systematically summarize the state-of-art fabrication techniques for implementing chip-scale integrated devices incorporating 2D materials, including approaches for material synthesis, on-chip transfer, film patterning, and property tuning / modification, as well as methods for the on-chip integration of 2D van der Waals heterostructures. We also discuss current challenges as well as the future outlook. While challenges remain, and more work is needed to improve the production scale and efficiency of these methods, there is no doubt that the on-chip integration of 2D materials will continue to underpin new key breakthroughs and greatly accelerate the practical applications of 2D materials beyond the laboratory given their atomic thickness representing the ultimate level of integration that can potentially be achieved.

**Acknowledgement**

This work was supported by the Australian Research Council Discovery Projects Programs (No. DP150102972, DP190103186, and FT210100806), the Swinburne ECR-SUPRA program, the Industrial Transformation Training Centers scheme (Grant No. IC180100005), the Beijing Natural Science Foundation (No. Z180007), the National Key R&D Program of China (2017YFA0303800), and the Chinese National Science Foundation (12134006).